\documentclass[sigconf]{acmart}
\usepackage{booktabs}  
\usepackage{array}    
\usepackage{tabularx}  % 자동 줄바꿈 X 열
\usepackage{graphicx}   % 이미지 포함
\usepackage{makecell}   % 셀 내 줄바꿈 간단히 할 때(아래 코드는 minipage로 처리)
\usepackage{multirow}
\usepackage{tabularx}
\usepackage{kotex}
\usepackage{caption}
\usepackage{url}
\usepackage{hyperref}
\usepackage{subcaption}
\usepackage[]{mdframed}
\usepackage{listings}
\usepackage{enumitem}
\usepackage{makecell}
\usepackage{adjustbox}
\usepackage[flushleft]{threeparttable}
\usepackage{tcolorbox}

\usepackage{stfloats}      % 또는 \usepackage{dblfloatfix}
\usepackage{placeins}      % 필요시 이동 제한용

\usepackage{colortbl}
\usepackage{xcolor}
\usepackage{multirow}
\usepackage{adjustbox}

 \AtBeginDocument{%
  \providecommand\BibTeX{{%
    \normalfont B\kern-0.5em{\scshape i\kern-0.25em b}\kern-0.8em\TeX}}}

\copyrightyear{2026}
\acmYear{2026}
\setcopyright{cc}
\setcctype{by}
\acmConference[WWW '26]{Proceedings of the ACM Web Conference 2026}{April 13--17, 2026}{Dubai, United Arab Emirates}
\acmBooktitle{Proceedings of the ACM Web Conference 2026 (WWW '26), April 13--17, 2026, Dubai, United Arab Emirates}
\acmPrice{}
\acmDOI{10.1145/3774904.3792728}
\acmISBN{979-8-4007-2307-0/2026/04}

\begin{document} 

\title{Moral Outrage Shapes Commitments Beyond Attention:\\Multimodal Moral Emotions on YouTube in Korea and the US}


\author{Seongchan Park}
\authornote{Equal contribution to this work.}
\affiliation{%
  % \institution{Korea Advanced Institute of Science and Technology}
  \institution{KAIST}
  \city{Daejeon}
  \country{Republic of Korea}
}
\email{sc.park@kaist.ac.kr}

\author{Jaehong Kim}
\authornotemark[1]
\affiliation{%
  % \institution{Korea Advanced Institute of Science and Technology}
  \institution{KAIST}
  \city{Daejeon}
  \country{Republic of Korea}
}
\email{luke.4.18@kaist.ac.kr}

\author{Hyeonseung Kim}
\affiliation{%
  % \institution{Korea Advanced Institute of Science and Technology}
  \institution{KAIST}
  \city{Daejeon}
  \country{Republic of Korea}
}
\email{kimyeonz@kaist.ac.kr}

\author{Heejin Bin}
\affiliation{%
  % \institution{Korea Advanced Institute of Science and Technology}
  \institution{KAIST}
  \city{Daejeon}
  \country{Republic of Korea}
}
\email{heejbin@kaist.ac.kr}

\author{Sue Moon}
\affiliation{%
  % \institution{Korea Advanced Institute of Science and Technology}
  \institution{KAIST}
  \city{Daejeon}
  \country{Republic of Korea}
}
\email{sbmoon@kaist.ac.kr}

\author{Wonjae Lee}
\affiliation{%
  % \institution{Korea Advanced Institute of Science and Technology}
  \institution{KAIST}
  \city{Daejeon}
  \country{Republic of Korea}
}
\email{wnjlee@kaist.ac.kr}

\begin{abstract}
Understanding how media rhetoric shapes audience engagement is crucial in the attention economy. This study examines how moral-emotional framing by mainstream news channels on YouTube influences user behavior across Korea and the United States. To capture the platform's multimodal nature, combining thumbnail images and video titles, we develop a multimodal moral emotion classifier by fine-tuning a vision–language model. The model is trained on human-annotated multimodal datasets in both languages and applied to approximately 400,000 videos from major news outlets. We analyze three engagement levels (views, likes, and comments), representing increasing degrees of commitment. The results show that \textit{other-condemning} rhetoric—expressions of moral outrage that criticize others’ morality—consistently increases all forms of engagement across cultures, with effect sizes strengthening from passive viewing to active commenting. These findings suggest that moral outrage is a particularly effective emotional strategy, attracting not only attention but also active participation. We discuss concerns about the potential misuse of \textit{other-condemning} rhetoric, as such practices may deepen polarization by reinforcing in-group/out-group divisions. To facilitate future research and ensure reproducibility, we publicly release our Korean and English multimodal moral emotion classifiers.\footnote{\url{https://github.com/Paul-scpark/Multimodal-Moral-Emotion}}
\end{abstract}




%                                                                                                \received[revised]{12 March 2009}


\maketitle
\vspace{-0.61em}

\newcommand\webconfavailabilityurl{https://zenodo.org/records/18368120}
\newcommand\webconfgithuburl{https://github.com/Paul-scpark/Multimodal-Moral-Emotion}

\ifdefempty{\webconfavailabilityurl}{}{
\begingroup\small\noindent\raggedright\textbf{Resource Availability:}\\
The source code used in this paper is publicly available via a Zenodo archive (DOI: \url{\webconfavailabilityurl}) and the corresponding GitHub repository (\url{\webconfgithuburl}).
\endgroup
}

\section{Introduction}

% Attention Economy와 도덕감정
Social media platforms operate within an attention economy, where human attention serves as the key currency of the digital marketplace~\cite {heitmayer2025second}. Because platforms monetize user engagement through advertising, they prioritize content that generates stronger interactions~\cite{van2024social, crockett2017moral}. Within this competitive environment, morally infused content stands out as a powerful driver of user participation, manifesting in diverse engagement behaviors such as likes, shares, and retweets~\cite{valenzuela2017behavioral, rathje2021out, brady2020attentional, dong2025parallel}. 
%The engagement effects of moral emotions are platform-dependent: \textit{other-condemning} rhetoric drives engagement on Reddit, whereas \textit{other-praising} content is more effective in dark web community~\cite{dong2025parallel}. This suggests that emotional impact depends not only on what moral emotions are expressed, but also on where and how they are communicated.
Prior research demonstrates that moral-emotional content spreads more widely, fosters civic participation, and intensifies polarization in online discourse~\cite{brady2017emotion, kim-etal-2024-moral, brady2025social}, while also accelerating the diffusion of misinformation~\cite{mcloughlin2024misinformation, solovev2022moral}. The prominence of moral emotions online is not incidental; it reflects a fundamental human tendency to be drawn to moral issues that structure social coordination and collective life~\cite{krebs2008morality,boehm2012moral}, highlighting their pivotal role in the digital attention economy.

% 도덕감정 연구는 NLP 분석에 국한, 멀티모달 충분히 반영 X
Despite growing evidence for the importance of moral emotions, existing studies leave two directions open for further exploration. First, most research has relied on unimodal, text-based approaches~\cite{brady2017emotion,brady2021social, solovev2022moral, kim-etal-2024-moral}, even though human communication and digital platforms are inherently multimodal, encompassing diverse modalities of observation and behavior~\cite{xu2023multimodal}. Building on this insight, multimodal analysis is essential for understanding how moral emotions are expressed in environments like YouTube, where images and text jointly evoke audience responses~\cite{yousefi2024examining}. Second, prior work has been heavily concentrated on Western contexts~\cite{brady2019ideological,hackenburg2023mapping, brady2020attentional, simchon2022troll}, raising questions about whether the dynamics of moral-emotional expression and engagement generalize across cultures. Earlier research in moral psychology has emphasized the significance of cross-cultural variation~\cite{haidt2003moral,malti2010development,van2024social}, underscoring the need for comparative analysis to build a more comprehensive understanding. 
% Examining moral emotions across distinct cultural settings is therefore crucial for assessing whether similar emotional cues elicit comparable engagement patterns among different audiences.

\begin{figure*}[th]
    \centering
    \includegraphics[width=0.93\linewidth]{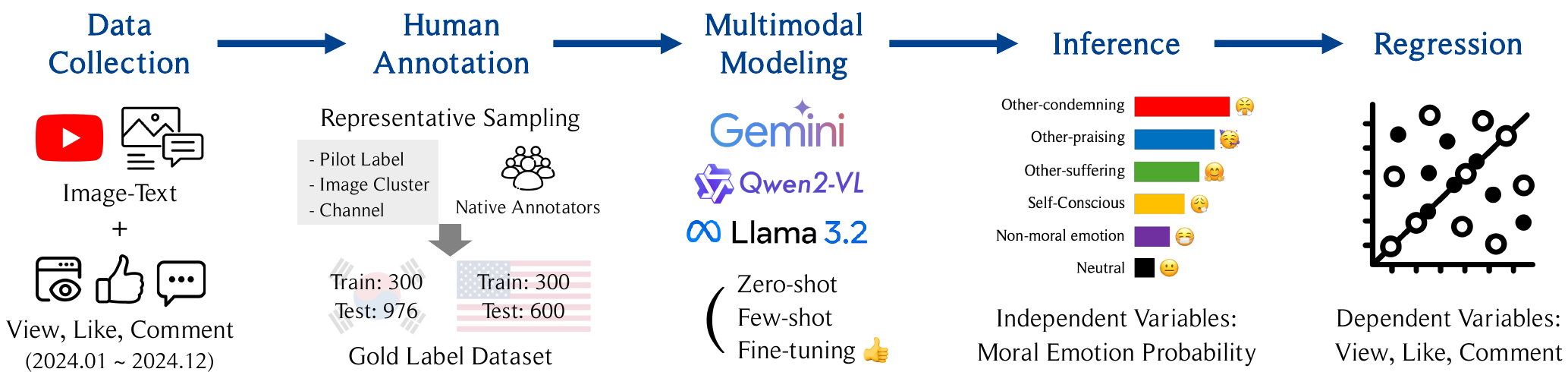}
    % \vspace{-1mm}
    \caption{Overview of the research framework. Multimodal YouTube data and human-labeled moral emotions are used to examine the relationship between predicted moral emotions and user engagement (views, likes, and comments).}
    \label{fig:process_overview}
\end{figure*}

% MLLMs 모델로 한국과 미국 주요 언론사 유튜브 콘텐츠 분석
% 1. MM 기반 도덕감정 분류 모델 구축
% 2. 한국과 미국 언론사 유튜브 콘텐츠로 도덕감정 표현 방식 비교
% 3. 회귀 분석으로 콘텐츠의 도덕감정이 User engagement에 미치는 영향과 문화적 차이 분석
To address these gaps, this study develops a multimodal framework for analyzing moral emotions in digital media environments, as illustrated in Figure~\ref{fig:process_overview}. We construct classification models using multimodal large language models (MLLMs) trained on human-annotated data of YouTube news content. Our dataset comprises approximately 400,000 videos collected between January and December 2024 from 7 Korean and 16 U.S. news channels, including thumbnail images, title texts, and associated metadata such as view, like, and comment counts.
% Our dataset consists of approximately 400,000 videos from 7 Korean and 16 U.S. news channels collected between January and December 2024, including thumbnails, titles, and engagement metadata.
% Our dataset comprises approximately 400,000 videos collected between January and December 2024 from 7 Korean and 16 U.S. news channels, including thumbnail images, title texts, and associated metadata such as view, like, and comment counts. 
To enable human annotation, content was sampled from the dataset by considering thumbnails and titles to capture multimodal cues. The annotation scheme consisted of four core moral emotions—\textit{other-condemning}, \textit{other-praising}, \textit{other-suffering}, and \textit{self-conscious}—as well as two additional categories: \textit{neutral} and \textit{non-moral emotion}, as defined in Table~\ref{tab:moral_emotion_descriptions}~\cite{haidt2003moral}. Following annotation by native speakers, items with majority agreement were retained, yielding 1,276 Korean and 900 English labeled samples. To build our classifiers, we apply zero-shot, few-shot, and fine-tuning approaches to open-source MLLMs, and evaluate closed-source models in zero-shot and few-shot settings. We then adopt the best-performing model as the final classifier, which is used to measure moral emotions across the entire dataset.

We analyze how moral emotions relate to three levels of user engagement (views, likes, and comments), each reflecting increasing degrees of commitment~\cite{kim2017like}. Our findings reveal a consistent cross-cultural pattern: \textit{other-condemning} rhetoric significantly increases all forms of engagement and is the only emotion that drives high-commitment behaviors such as liking and commenting. Its effect grows progressively across engagement levels, with the strongest impact observed for comments. These results suggest that \textit{other-condemning} rhetoric is a highly effective strategy for capturing attention and motivating participation. We discuss how its emotionally charged nature, which inherently divides audiences into in-groups and out-groups, carries the risk of deepening
polarization when amplified or misused~\cite{brady2020mad,solovev2022moral,simchon2022troll}.
\section{Related Work}
\subsection{Moral Emotion and User Engagement}
Humans are evolutionarily drawn to moral issues~\cite{boehm2012moral,krebs2008morality}, and this tendency is amplified within the attention economy of social media~\cite{van2024social}. Prior work shows that content expressing moral emotions attracts higher user engagement and spreads more widely online~\cite{brady2017emotion}. Among these emotions, \textit{other-condemning} rhetoric is particularly influential: people experience stronger \textit{other-condemning} (moral outrage) responses when encountering immoral acts online compared to offline contexts~\cite{crockett2017moral}, and such rhetoric boosts sharing even when the content contains misinformation~\cite{solovev2022moral,mcloughlin2024misinformation}. 
Beyond attention and information diffusion, the influence of moral outrage also extends to broader forms of collective behavior.
% Importantly, the influence of moral outrage extends beyond attention and information diffusion.
Research on government-led online petitions demonstrates that moral emotions shape civic participation, where \textit{other-condemning} rhetoric fostered polarization in both Korea and the United Kingdom~\cite{kim-etal-2024-moral}. Building on these literature, we examine how moral-emotional rhetoric in news content on YouTube—where visual and textual cues shape perception—drives audience engagement across viewing, liking, and commenting behaviors on social media. 

\subsection{Measurement of Moral Emotion}
Approaches to measuring moral emotions have evolved alongside advances in natural language processing, shifting from traditional lexicon-based techniques to word embeddings and transformer architectures. Early studies extracted overlapping terms from moral foundation and emotion lexicons to construct moral-emotional word sets~\cite{brady2017emotion,solovev2022moral}. Subsequently, embedding-based models enabled the detection of specific affective states such as \textit{other-condemning}~\cite{brady2021social}. More recently, large-scale annotated datasets have supported transformer-based classifiers that cover a full range of moral emotions, including \textit{other-suffering}, \textit{other-praising}, and \textit{self-conscious}~\cite{kim-etal-2024-moral}. These models, trained on parallel Korean and English annotations, provide the first resources for measuring moral emotions and thereby addressing calls for cross-cultural analyses of morality in online discourse~\cite{van2024social}. Moving beyond predominantly unimodal approaches, we propose a multimodal classifier that measures moral emotions expressed through both images and text across cultural contexts.

\section{Multimodal YouTube Dataset}
\subsection{Data Collection}
% KOR: (MBC, SBS, KBS, YTN, JTBC, Channel A, TV Chosun)
% US: (MSNBC, ABC News, AP, CBS News, CNN, NBC News, Newsnation, Wall Street Journal, New York Post, The Washington Times, Breitbart, Blaze Media, CBN, Fox News, Newsmax, and OAN)

We collected a comprehensive dataset of YouTube videos uploaded between January and December, 2024, from the official channels of major news sources in Korea and the U.S. To ensure a representative sample, we included 7 Korean and 16 American channels, chosen based on subscriber size and political diversity. The political leanings of Korean outlets were identified from~\citet{hahn2015fragmentation}, while their U.S. counterparts were determined by referencing the AllSides Media Bias Chart.\footnote{\url{https://www.allsides.com/media-bias/media-bias-chart}} Further details on the selected channels, including their creation dates, subscriber counts, number of uploaded videos, and total views, are summarized in Appendix Table~\ref{tab:channels_info}. The period covers two major political events: the Korean parliamentary election in April and the U.S. presidential election in November.

For each video, our dataset contains thumbnail images and title texts as the primary multimodal inputs, along with metadata such as video descriptions, IDs, durations, upload dates, and engagement indicators including view, like, and comment counts. To ensure comparability across videos with different upload dates, all metadata were uniformly retrieved during a fixed eight-day window from February 5 to 13, 2025. This approach mitigates potential bias in engagement metrics that could arise from varying view accumulation periods. To validate this methodological choice, we tracked the daily view counts of 1,703 videos for 60 days post-release and confirmed that view count growth stabilizes over time, with the average daily growth rate declining to below 0.04\% after 60 days. A detailed description and visualization of this validation process are provided in Appendix~\ref{sec:validation_data_collection}. Data collection was conducted using a combination of the official YouTube Data API\footnote{\url{https://developers.google.com/youtube/v3}} and the open-source package yt-dlp\footnote{\url{https://github.com/yt-dlp/yt-dlp}}, which together enabled reliable large-scale scraping of both video- and comment-level information. In total, our dataset comprises 397,897 videos and their associated user responses (292,136 from Korea; 105,761 from the United States), facilitating a robust foundation for multimodal moral emotion classification and cross-cultural analysis of engagement dynamics.

\begin{table*}[!t]
\centering
\small
\renewcommand{\arraystretch}{1.3}
\caption{Descriptive statistics of engagement metrics across countries.}
\begin{adjustbox}{max width=\textwidth}
\begin{tabular}{lccccccccccccccc}
\toprule
\multirow{2}{*}{\textbf{Metric}} &
\multicolumn{7}{c}{\textbf{Korea (\textit{N}=292,136)}} &
\multicolumn{7}{c}{\textbf{United States (\textit{N}=105,761)}} \\
\cmidrule(lr){2-8} \cmidrule(lr){9-15}
 & Mean & Median & SD & Min & Max & Skew & Kurt 
 & Mean & Median & SD & Min & Max & Skew & Kurt \\
\midrule
View     & 41031.44 & 4129.00 & 158107.57 & 4.00 & 11472122.00 & 12.90 & 343.05
         & 92134.26 & 16361.00 & 244354.79 & 1.00 & 13123214.00 & 10.39 & 242.44 \\
Like     & 525.72 & 52.00 & 1998.48 & 0.00 & 77035.00 & 10.96 & 193.59
         & 1593.85 & 258.00 & 4161.66 & 0.00 & 169322.00 & 8.73 & 150.24 \\
Comment  & 214.32 & 21.00 & 772.01 & 0.00 & 42081.00 & 11.75 & 255.87
         & 585.28 & 121.00 & 1411.53 & 0.00 & 137601.00 & 15.59 & 947.91 \\
\bottomrule
\multicolumn{15}{r}{\footnotesize Notes: SD = standard deviation; Skew = skewness; Kurt = kurtosis} \\
\end{tabular}
\end{adjustbox}
\label{tab:engagement_stats}
\end{table*}

\subsection{Data Description}
\label{sec:data_description}

This section offers a comprehensive overview of the collected dataset, describing key patterns of user behavior and the main topics covered in the videos. Table~\ref{tab:engagement_stats} summarizes the descriptive statistics of engagement metrics—views, likes, and comments—for both the Korean and U.S. datasets, revealing the heavy-tailed nature of online attention distributions~\cite{lin2011more}. 

On average, U.S. news videos received 92,134 views, more than twice the 41,031 views observed for Korean videos. Both datasets exhibit strong long-tail patterns, but the median-to-mean ratio for Korea (0.10) is lower than that for the U.S. (0.18), indicating a higher degree of inequality in view distribution among Korean videos. Regarding user reactions, likes also follow a heavy-tailed distribution, with U.S. videos showing higher average counts but a comparable level of relative dispersion across both regions. The comment intensity, measured as the ratio of comments to views, is likewise similar between the two datasets (0.52\% in Korea; 0.64\% in the U.S.). Although the overall commenting rate differs only marginally, both regions display substantial engagement concentration, with skewness values of 11.75 (Korea) and 15.59 (U.S.), respectively. The consistently high skewness and kurtosis across all metrics confirm that a small subset of viral videos captures the majority of user interactions, reflecting the unequal distribution of attention that typifies online news ecosystems~\cite{lerman2010using}.

To further characterize the content landscape of our dataset, we applied a multimodal BERTopic framework to uncover the major themes embedded in the collected videos~\cite{grootendorst2022bertopic}. This approach enabled us to identify the topical structures within each media ecosystem and analyze which thematic clusters exhibit stronger associations with particular moral emotions. We extracted semantic representations from both video thumbnails and titles using the CLIP model~\cite{radford2021learning}—specifically \texttt{clip-vit-large-patch14-ko}\footnote{\url{https://huggingface.co/Bingsu/clip-vit-large-patch14-ko}} for Korean and \texttt{clip-vit-base-patch32}\footnote{\url{https://huggingface.co/openai/clip-vit-base-patch32}} for English. The resulting high-dimensional embeddings were reduced using UMAP and clustered into distinct topics using the HDBSCAN~\cite{mcinnes2018umap, mcinnes2017hdbscan}. 

\begin{figure*}[ht]
    \centering
    \includegraphics[width=1\linewidth]{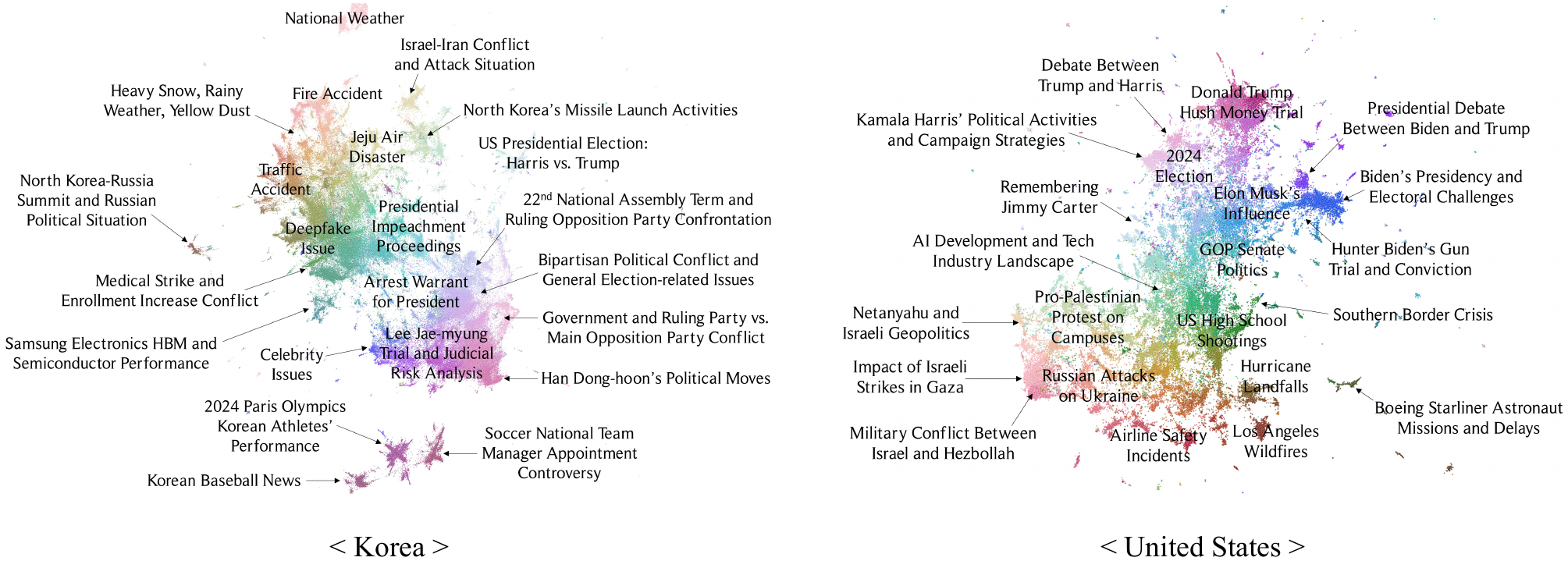}
    \vspace{-5mm}
    \caption{Topic distributions derived from BERTopic for Korean and English YouTube news data (January–December 2024).}
    % Korean topic labels were translated into English for interpretability.
    \label{fig:topic_modeling}
\end{figure*}

To assign human-interpretable labels to the extracted topics, we adopted an ensemble approach using \texttt{gemini-2.5-flash-lite} and \texttt{gpt-4o-mini}. Before label generation, we removed common or non-informative terms such as broadcaster names and generic stopwords to refine keyword quality. For each topic, the models supplied with 9 representative video samples—selected based on c-TF-IDF scores and consisting of both thumbnails and titles—along with keywords extracted via Maximal Marginal Relevance (MMR). MMR was applied with a diversity parameter of 0.5 to generate more varied keywords and minimize redundancy across semantically similar terms. The two language models independently generated candidate topic labels, which were then compared using cosine similarity. Labels with similarity scores of 0.9 or higher were automatically accepted, while those below the threshold underwent manual review by researchers to ensure accuracy and coherence. After a final curation step that filtered out semantically incoherent or irrelevant clusters, we obtained 123 topics for the Korean and 169 topics for the U.S. dataset. Figure~\ref{fig:topic_modeling} visualizes the overall topic distributions, highlighting distinct thematic compositions between the Korean and U.S. news domains.

\section{Method}

\subsection{Human Annotation}

Since, to the best of our knowledge, no multimodal resource on moral emotions exists, we constructed a new annotated dataset. To obtain a representative sample, we employed a multi-stage sampling strategy. First, pilot labels for the six moral emotion categories were assigned to video titles using a state-of-the-art text classification model for moral emotions~\cite{kim-etal-2024-moral}. Second, we used BERTopic with the \texttt{CLIP-ViT-B-32}\footnote{\url{https://huggingface.co/sentence-transformers/clip-ViT-B-32}} embedding model to cluster thumbnails into groups of visually similar images~\cite{grootendorst2022bertopic}. Finally, text-based pilot labels, image-based clusters, and channel distribution were jointly considered to sample 1,500 thumbnail–title pairs for each language domain, ensuring both balanced coverage of moral emotion categories and diversity across news sources.

Annotations were conducted in two rounds of 750 samples each. Each round involved three raters, resulting in six participants per language domain. All annotators were native speakers who regularly consume YouTube as one of their main social media platforms. They received detailed guidelines based on the definitions of moral emotions in \citet{haidt2003moral}, accompanied by illustrative examples. For each thumbnail–title pair, raters were instructed to identify the expressed moral emotion from six predefined categories, with the option to select \textit{Hard to tell} if none applied. Gold labels were determined by majority vote, and pairs without a majority decision were excluded from the final dataset. The full annotation guideline and interface are provided in Appendix~\ref{appendix:annotation_guideline}.

\begin{table}[hb]
    \centering
    \caption{The inter-annotator agreement (IAA) score of human-annotated dataset.} 
    \begin{tabular}{l|c|c}
    \toprule
    \textbf{Metric} & \textbf{Korean} & \textbf{English} \\ \midrule
    Cohen's kappa & 0.7514 & 0.5278 \\
    Fleiss' kappa & 0.7512 & 0.5271 \\
    Krippendorff's alpha & 0.7512 & 0.5272 \\
    \bottomrule
    \end{tabular}
    \label{tab:agreement-score}
    % \vspace{-2mm}
\end{table}

\begin{table*}[]
\caption{Model performance on Korean (KO) and English (EN) datasets across closed-source and open-source models. Results are shown for zero-shot (ZS), few-shot (FS), and fine-tuning (FT) settings, with accuracy (Acc.) and macro F1 scores reported for each moral emotion category and the overall average. Best performance for each category is marked with bold.}
\resizebox{\textwidth}{!}{
\begin{tabular}{@{}ccc|cccccccccccc|cc@{}}
\toprule
\multirow{2}{*}{\textbf{Lang}} & \multirow{2}{*}{\textbf{Model}} & \multirow{2}{*}{\textbf{Method}} 
& \multicolumn{2}{c}{\textbf{Other-condemning}}
& \multicolumn{2}{c}{\textbf{Other-praising}}
& \multicolumn{2}{c}{\textbf{Other-suffering}} 
& \multicolumn{2}{c}{\textbf{Self-conscious}} 
& \multicolumn{2}{c}{\textbf{Neutral}}
& \multicolumn{2}{c|}{\textbf{Non-moral emotion}} 
& \multicolumn{2}{c}{\textbf{Avg}} \\ 
\cmidrule(lr){4-17}
 &  &  
& Acc. & F1 & Acc. & F1 & Acc. & F1 & Acc. & F1 & Acc. & F1 & Acc. & F1 & Acc. & F1 \\ 
\midrule
\multirow{4}{*}{KO} 
 & \multirow{2}{*}{\makecell[c]{Closed-source\\(Gemini-2.5-Flash-Lite)}} & ZS & 0.8156 & 0.7741 & \textbf{0.9375} & \textbf{0.8851} & 0.7930 & 0.7027 & 0.9283 & \textbf{0.8945} & 0.8842 & 0.8055 & 0.3648 & 0.3514 & 0.7872 & 0.7356 \\
 &   & FS & 0.7643 & 0.7258 & 0.9293 & 0.8763 & 0.7510 & 0.6667 & 0.6434 & 0.6188 & 0.5502 & 0.5370 & 0.3719 & 0.3516 & 0.6684 & 0.6293 \\ 
\cmidrule(lr){2-17}
 & \multirow{3}{*}{\makecell[c]{Open-source\\(Qwen2-VL-7B-Instruct)}} & ZS & 0.7223 & 0.5140 & 0.7982 & 0.5013 & 0.8084 & 0.5152 & 0.7295 & 0.5209 & 0.6844 & 0.4906 & 0.7725 & 0.4942 & 0.7526 & 0.5061 \\
 &   & FS & 0.7469 & 0.5748 & 0.8514 & 0.4599 & 0.7807 & 0.5313 & 0.7305 & 0.5569 & 0.7941 & 0.4475 & 0.7941 & 0.5276 & 0.7830 & 0.5163 \\
 &   & FT & \textbf{0.8801} & \textbf{0.8012} & 0.9344 & 0.8719 & \textbf{0.9016} & \textbf{0.7856} & \textbf{0.9293} & 0.8875 & \textbf{0.9016} & \textbf{0.8431} & \textbf{0.8412} & \textbf{0.6166} & \textbf{0.8981} & \textbf{0.8010} \\ 
\midrule
\multirow{4}{*}{EN} 
 & \multirow{2}{*}{\makecell[c]{Closed-source\\(Gemini-2.5-Flash-Lite)}} & ZS & \textbf{0.8283} & \textbf{0.7642} & 0.8517 & \textbf{0.6808} & \textbf{0.8000} & \textbf{0.7219} & \textbf{0.9317} & \textbf{0.6931} & 0.7850 & \textbf{0.6824} & 0.5583 & 0.5108 & 0.7925 & \textbf{0.6755} \\
 &   & FS & 0.7983 & 0.7329 & 0.8567 & 0.6797 & 0.7650 & 0.7048 & 0.9100 & 0.6689 & \textbf{0.7983} & 0.6441 & 0.4167 & 0.4104 & 0.7575 & 0.6401 \\ 
\cmidrule(lr){2-17}
 & \multirow{3}{*}{\makecell[c]{Open-source\\(Qwen2-VL-7B-Instruct)}} & ZS & 0.7917 & 0.4496 & 0.8283 & 0.4803 & 0.6783 & 0.4827 & 0.8683 & 0.5206 & 0.7367 & 0.4638 & \textbf{0.6583} & 0.4775 & 0.7603 & 0.4791 \\
 &   & FS & 0.7817 & 0.4603 & \textbf{0.8600} & 0.4624 & 0.7800 & 0.4382 & 0.9100 & 0.5108 & 0.7250 & 0.5743 & 0.6267 & 0.4931 & 0.7806 & 0.4898 \\
 &   & FT & 0.8250 & 0.7240 & \textbf{0.8600} & 0.6686 & 0.7900 & 0.7036 & 0.8717 & 0.5422 & 0.7917 & 0.6141 & 0.6567 & \textbf{0.5338} & \textbf{0.7992} & 0.6310 \\ 
\bottomrule
\end{tabular}}
\label{tab:performance}
\end{table*}

The final human-annotated dataset consists of 1,276 Korean and 900 U.S. thumbnail–title pairs. Inter-annotator agreement, measured by Cohen’s kappa, Fleiss’ kappa, and Krippendorff’s alpha~\cite{cohen1960coefficient, fleiss1971measuring, krippendorff1970estimating}, averaged 0.7513 for Korean and 0.5274 for English (Table~\ref{tab:agreement-score}). The higher agreement in Korean data may be attributed to the frequent inclusion of captions in thumbnails and the media’s strong reliance on visual and textual cues, which made the underlying moral-emotional rhetoric more explicit. Representative pairs for each category in both languages are provided in Appendix Table~\ref{tab:moral_examples_split}.

\subsection{Modeling}

\noindent\textbf{Human Gold Label Split:} To develop multimodal moral emotion classifiers, we began with the human-annotated dataset, comprising 1,276 Korean and 900 English samples. Given the high quality but limited size of this dataset, our strategy was to leverage pre-trained MLLMs known for their strong performance on emotion-related tasks~\cite{liu2025mmaffben}. For consistent evaluation across various modeling approaches (e.g., zero-shot, few-shot, and fine-tuning), each language dataset was divided into training and test subsets. A fixed set of 300 samples was designated for training, serving both as fine-tuning data and as a pool of examples for few-shot prompting, while the remaining samples formed a held-out test set for uniform performance evaluation. To maintain the proportional distribution of the six moral emotion categories across splits, stratified sampling was applied, resulting in 300/976 samples for Korean and 300/600 for English in the training/test sets.

\vspace{0.5em}
\noindent\textbf{Closed- and Open-source MLLM:} Our modeling approach employed both closed- and open-source MLLMs capable of processing multimodal inputs. We framed the task as a binary classification for each of the six moral emotions rather than a single multi-class setting. Preliminary experiments showed that independent binary classifiers performed more robustly, allowing each model to capture the specific nuances of a single target emotion without interference from others. Consequently, we developed 12 distinct models in total, one per emotion for both the Korean and English.

As a baseline, we evaluated the closed-source model \texttt{Gemini 2.5 Flash-Lite} using zero-shot and few-shot prompting. For the few-shot setup, six representative examples (one per moral emotion) were drawn from the training data and incorporated into the prompt. The model was instructed to respond with either \textit{True} or \textit{False} to indicate whether a given thumbnail–title pair expressed a specific moral emotion. Each prompt, presented in the native language of the dataset, included a definition of the target emotion (an example is shown in Appendix Figure~\ref{fig:prompt_structure}). Overall, the zero-shot configuration slightly outperformed the few-shot method, achieving average accuracies of 0.7872 (macro F1=0.7356) for Korean and 0.7925 (F1=0.6755) for English.

For the open-source approach, we used \texttt{Qwen2-VL-7B-Instruct} and evaluated it under the same zero-shot and few-shot setups as the closed-source model, with an additional fine-tuning stage~\cite{Qwen2VL}. Fine-tuning on the 300 training samples aimed to enhance the model’s contextual understanding of moral emotions beyond its pre-trained knowledge. Since the base model was originally designed for text generation, we adapted its architecture for binary classification by resizing the final output layer to a single dimension. This allowed the model to produce a logit for each input, which was then converted into a probability score indicating the presence of the target emotion. The model was trained using an SGD optimizer with a learning rate of 1e-5, a momentum of 0.9, and weight decay of 1e-5, with Binary Cross Entropy as the loss function. As expected, fine-tuning yielded the highest performance, with accuracies of 0.8981 (F1=0.8010) for the Korean and 0.7992 (F1=0.6310) for the English dataset. For comparison, we also fine-tuned a \texttt{Llama-3.2-11B-Vision-Instruct} model under the same configuration, but Qwen consistently performed better; detailed results are provided in Appendix~\ref{appendix:qwen_llama_comparison}.

\vspace{0.5em}
\noindent\textbf{Model Comparison and Analysis:} A comprehensive comparison of all model performances is detailed in Table~\ref{tab:performance}. Fine-tuning consistently achieved the best overall results, particularly for the Korean dataset, where it reached the highest average accuracy (0.8981) and F1 score (0.8010). In the English dataset, it also secured the best accuracy (0.7992), while the closed-source model in zero-shot mode obtained a slightly higher F1 score (0.6755). However, the reliance on paid APIs in closed-source systems imposes practical limitations for large-scale applications.

To benchmark the contribution of multimodality, we compared these results against a text-only baseline using a BERT-based classifier that received only video titles as input~\cite{kim-etal-2024-moral}. The text-only model achieved accuracies of 0.7500 (F1=0.7324) for Korean and 0.4950 (F1=0.4730) for English. The substantial performance gap between our fine-tuned model and the text-only baseline strongly validates our approach. These findings confirm that incorporating both visual and textual cues is essential for classifying moral emotions in multimodal environments such as YouTube.

\subsection{Inference}

\begin{figure}[h]
    \vspace{-1mm}
    \centering
    \includegraphics[width=1\linewidth]{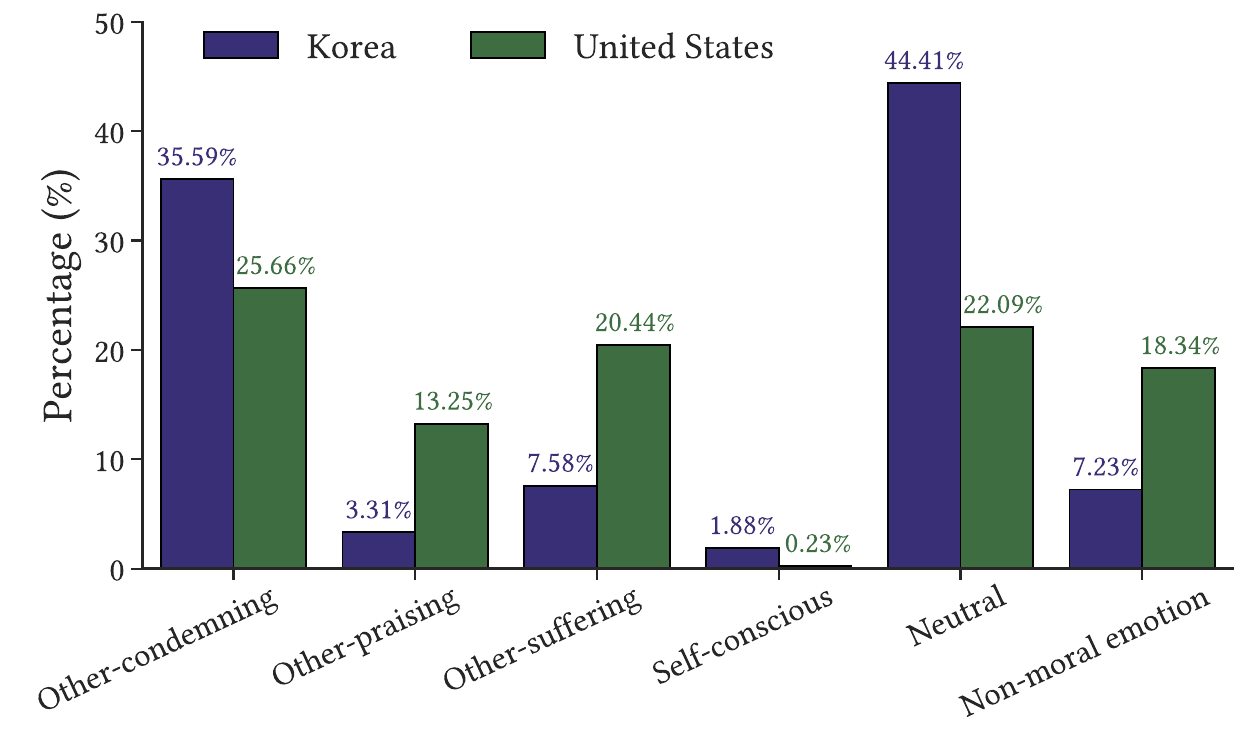}
    \vspace{-7mm}
    \caption{Distribution of primary moral emotions in the Korean and U.S. datasets, categorized by the highest predicted probability from the fine-tuned models.}
    \label{fig:inference_count}
\end{figure}

\vspace{-1mm}

To analyze the overall landscape of moral emotions in YouTube news content, we applied our 12 fine-tuned binary classifiers to the full dataset, comprising 292,136 Korean and 105,761 U.S. videos. For each video, the six classifiers independently calculated the probability (a value between 0 and 1) of their respective moral emotion being present. Figure~\ref{fig:inference_count} illustrates the proportional distribution of primary moral emotions, where each video is categorized by the emotion with the highest predicted probability. In the Korean dataset, the majority of content was classified as \textit{neutral} (44.41\%), followed by \textit{other-condemning} (35.59\%), \textit{other-suffering}, \textit{non-moral emotion}, \textit{other-praising}, and \textit{self-conscious}. The English dataset exhibited a similar trend, with \textit{other-condemning} (25.66\%) and \textit{neutral} (22.09\%) being the most prevalent. These results highlight that across both countries, moral condemnation and neutral-toned content dominate YouTube’s news media ecosystems.

\begin{table*}[]
\caption{Top three topics with the highest predicted probabilities for each moral emotion category. Probabilities represent the average predicted values from our multimodal fine-tuned model across all samples within each emotion.}
\resizebox{\textwidth}{!}{
\begin{tabular}{@{}ccccc@{}}
\toprule
\multicolumn{2}{l}{} & \multicolumn{2}{c}{\textbf{Korea}} & \multicolumn{1}{c}{\textbf{United States}} \\ \midrule
\multicolumn{2}{c}{Other-condemning} & \multicolumn{2}{c}{\begin{tabular}[c]{@{}c@{}}Political Crisis of the Yoon Suk Yeol Administration\\ Administration's Probes \& Political Controversies\\ Medical Student Girlfriend Murder Case\end{tabular}} & \begin{tabular}[c]{@{}c@{}}Donald Trump Hush Money Trial\\ Carl Higbie's Political Commentary\\ Conservative Media Criticism of Kamala Harris\end{tabular} \\ \midrule
\multicolumn{2}{c}{Other-praising} & \multicolumn{2}{c}{\begin{tabular}[c]{@{}c@{}}2024 Paris Olympics Korean Athletes' Performance\\ Nobel Prize in Literature for Han Kang\\ Korean Baseball News\end{tabular}} & \begin{tabular}[c]{@{}c@{}}Remembering Jimmy Carter\\ Jimmy Carter's Funeral Events\\ WWII Anniversaries and Veteran Tributes\end{tabular} \\ \midrule
\multicolumn{2}{c}{Other-suffering} & \multicolumn{2}{c}{\begin{tabular}[c]{@{}c@{}}Heavy Rain and Flood Damage\\ Fishing Vessel Accident Casualties\\ LA Wildfire Spread and Damage Situation\end{tabular}} & \begin{tabular}[c]{@{}c@{}}Deadly Floods And Landslides\\ Impact of Israeli Strikes in Gaza\\ Deadly Taiwan Earthquake\end{tabular} \\ \midrule
\multicolumn{2}{c}{Neutral} & \multicolumn{2}{c}{\begin{tabular}[c]{@{}c@{}}National Weather\\ Heavy Rainfall Damage\\ Fluctuating Weather \& Temperatures\end{tabular}} & \begin{tabular}[c]{@{}c@{}}Ivory Hecker Tonight: Political Commentary\\ Israel-Hamas Conflict and Tensions\\ Lawrence O'Donnell's Commentary on Donald Trump\end{tabular} \\ \bottomrule
\end{tabular}}
\vspace{2mm}
\label{tab:topic_examples}
\end{table*}

To qualitatively validate the model’s predictions, we analyzed the relationship between the predicted emotion probabilities and the thematic topics identified in Section~\ref{sec:data_description}. 
Table~\ref{tab:topic_examples} shows the top three topics with the highest average probabilities for each moral emotion category (excluding low-performing ones such as \textit{self-conscious} and \textit{non-moral emotion}), revealing a clear and intuitive alignment. For instance, \textit{other-condemning} content was predominantly associated with political scandals and criminal justice, while \textit{other-praising} centered on achievements and commemorative events. Likewise, \textit{other-suffering} content corresponded to natural disasters and accidents. This alignment between emotional classification and topical content provides qualitative evidence that our fine-tuned model effectively captures and distinguishes moral-emotional expressions in real-world media contexts.

\section{Results}
\subsection{Regression Model Specification}
We estimated three engagement-related count variables (i.e., view, like, and comment counts) using negative binomial regression models, with emotion variables and control covariates for both YouTube datasets. This approach was justified by data overdispersion—where variance exceeded the mean—and all overdispersion parameters were significant at the $\alpha=.01$ level. For emotion predictors, we used probability scores (0–1) from the fine-tuned classifier for four categories (\textit{other-condemning}, \textit{other-praising}, \textit{other-suffering}, \textit{neutral}) as continuous variables to capture mixed emotional signals. The \textit{self-conscious} and \textit{non-moral emotion} categories were excluded due to their low F1 scores (below 0.6) in English data, following prior research practice~\cite{kim-etal-2024-moral}. To account for differences in video duration, baseline audience size across channels, and temporal variations such as month and weekday effects, we included these variables as controls in all models.

\subsection{Regression Result}

\begin{table}[t!]
\caption{Estimated coefficients from negative binomial regression models by country and engagement outcome. All models include fixed effects for video duration, channel, month, and weekday. \textit{Significance levels:} $^{*}p<.05$, $^{**}p<.01$, $^{***}p<.001$ (\colorbox{green!25}{positive}, \colorbox{red!25}{negative}).}
\vspace{-1mm}
\centering
\small
\frenchspacing
\renewcommand{\arraystretch}{1.25}
\resizebox{1\columnwidth}{!}{
\begin{tabular}{c|ccc|ccc}
\toprule
& \multicolumn{3}{c|}{\textbf{Korea (\textit{N}=292,136)}} & \multicolumn{3}{c}{\textbf{United States (\textit{N}=105,761)}} \\
\cmidrule(lr){2-7}
& \textbf{View} & \textbf{Like} & \textbf{Comment} & \textbf{View} & \textbf{Like} & \textbf{Comment} \\
\midrule
Other-condemning & \cellcolor{green!25}0.035$^{***}$ & \cellcolor{green!25}0.151$^{***}$ & \cellcolor{green!25}0.547$^{***}$ & \cellcolor{green!25}0.673$^{***}$ & \cellcolor{green!25}0.834$^{***}$ & \cellcolor{green!25}1.124$^{***}$ \\
Other-praising   & \cellcolor{red!25}-0.526$^{***}$ & \cellcolor{green!25}0.002 & \cellcolor{red!25}-0.952$^{***}$ & \cellcolor{red!25}-0.353$^{***}$ & \cellcolor{red!25}-0.061$^{***}$ & \cellcolor{red!25}-0.177$^{***}$ \\
Other-suffering  & \cellcolor{green!25}0.141$^{***}$ & \cellcolor{red!25}-0.192$^{***}$ & \cellcolor{red!25}-0.367$^{***}$ & \cellcolor{red!25}-0.084$^{***}$ & \cellcolor{red!25}-0.300$^{***}$ & \cellcolor{red!25}-0.538$^{***}$ \\
Neutral          & \cellcolor{red!25}-1.306$^{***}$ & \cellcolor{red!25}-1.296$^{***}$ & \cellcolor{red!25}-1.505$^{***}$ & \cellcolor{green!25}0.130$^{***}$ & \cellcolor{red!25}-0.105$^{***}$ & \cellcolor{red!25}-0.632$^{***}$ \\
% \midrule
% \textit{Controls} & & & & & & \\
% \quad Duration & \multicolumn{3}{c|}{Yes} & \multicolumn{3}{c}{Yes} \\
% \quad Channel Fixed Effects & \multicolumn{3}{c|}{Yes} & \multicolumn{3}{c}{Yes} \\
% \quad Month Fixed Effects & \multicolumn{3}{c|}{Yes} & \multicolumn{3}{c}{Yes} \\
% \quad Weekday Fixed Effects & \multicolumn{3}{c|}{Yes} & \multicolumn{3}{c}{Yes} \\
\bottomrule
\end{tabular}
}
\label{tab:country_emotion_coef}
% \vspace{-2mm}
\end{table}

\begin{figure*}[th]
    \centering
    \includegraphics[width=0.8\linewidth]{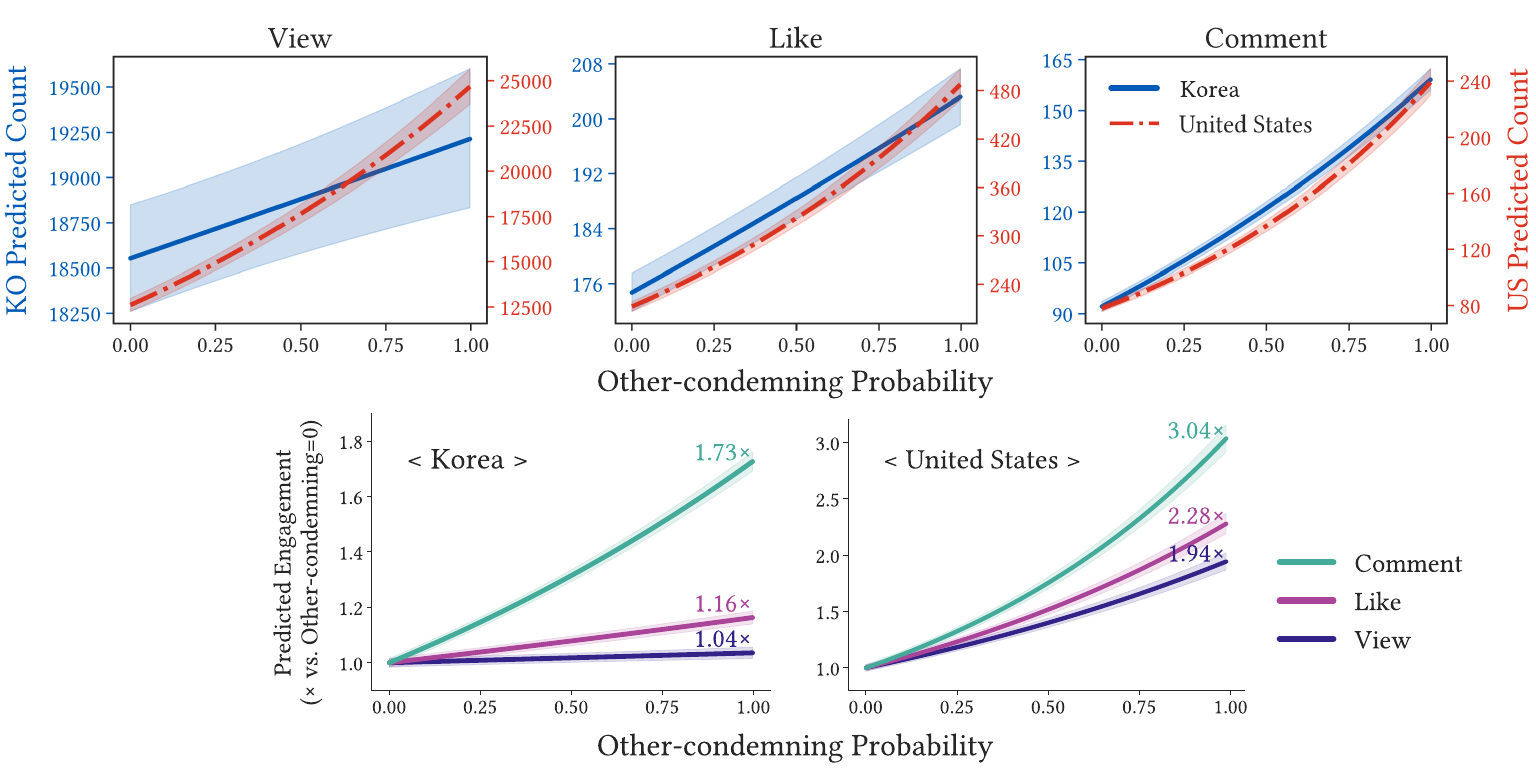}
    \vspace{-1mm}
    \caption{
Predicted engagement by \textit{other-condemning} emotion probability. 
Top panels show fitted counts from the negative binomial models, and bottom panels show relative engagement (IRR) for Korea and the United States. 
As moral outrage intensity increases, engagement progressively rises from views to comments. 
Shaded areas denote 95\% confidence intervals.
}
    \label{fig:regression}
\end{figure*}

Table~\ref{tab:country_emotion_coef} summarizes the effects of moral-emotional rhetoric on user engagement after adjusting for control variables. 
% Color shading indicates the direction of association—green for positive and red for negative coefficients. 
Across both countries, \textit{other-condemning} rhetoric—expressions of moral outrage—was consistently linked to higher audience engagement. In Korea, content with strong moral condemnation generated about 4\% higher views, 16\% more likes, and 73\% greater comment activity compared to content without such rhetoric (IRR=1.04, 1.16, and 1.73). The pattern was even stronger in the United States, where engagement nearly doubled for views, more than doubled for likes, and tripled for comments (IRR=1.94, 2.28, and 3.04). These results indicate that moral outrage not only attracts attention but also drives deeper participation, from passive viewing to active engagement, such as liking and commenting. Because the predictors are probabilities ranging from 0 to 1, these IRR values represent the upper-bound effects of emotional intensity. Figure~\ref{fig:regression} visually illustrates this pattern: as the probability of \textit{other-condemning} rhetoric increases, predicted engagement rises steadily across all behavioral levels. To rule out sample- or channel-specific effects, we conducted robustness checks using bootstrap resampling and channel-level regressions, which yielded consistent positive effects of \textit{other-condemning} rhetoric in both countries (see Appendix~\ref{sec:regression_robustness_analysis} for details).

% To assess the robustness of these findings, we conducted additional validation analyses using bootstrap resampling and channel-level regressions. The positive association between other-condemning rhetoric and engagement remained stable across 1,000 bootstrap subsamples and was consistently observed across individual news channels in both countries (see Appendix~A.5).

% \pagebreak

Other predictors, by contrast, showed negative or inconsistent associations with engagement. \textit{Other-praising} rhetoric decreased views and comments in Korea and consistently reduced all three engagement metrics in the U.S., suggesting that moral praise tends to suppress rather than stimulate online interaction. The \textit{other-suffering} emotion increased attention-level engagement (views) in Korea but decreased higher-commitment behaviors such as likes and comments. In the United States, it showed consistently negative associations across all engagement metrics.
\textit{Neutral} decreased all engagement metrics in Korea, whereas in the U.S. it increased view counts but reduced likes and comments.
Taken together, except for \textit{other-condemning} rhetoric, no other emotional category fostered more committed engagement such as likes or comments.

\section{Discussion}
Methodologically, this study developed a multimodal moral emotion classification model applicable to both Korean and English contexts, addressing an urgent gap in the moral emotion literature
~\cite{van2024social}. By incorporating visual (thumbnail) and textual (title) cues, our approach better captures the perceptual environment in which users decide whether to engage with YouTube content. We further examined how different moral emotions shape different forms of user engagement—views, likes, and comments—reflecting increasing degrees of behavioral commitment. Through this methodological advancement, we derived theoretical insights into how moral outrage functions within the attention economy.

We found that across both countries, only \textit{other-condemning} rhetoric predicted audience engagement beyond mere attention.
The effect grew stronger for actions requiring more commitment (from viewing to liking to commenting), suggesting that outrage serves as a powerful motivator in digital communication. This pattern supports prior work showing that moral-emotional content captures attention and fosters participation~\cite{brady2017emotion}, but extends it by demonstrating that outrage also drives commitment-level in environments where images and text jointly shape user engagement.

However, the same mechanism that enhances engagement may also foster polarization, especially when the emotional nature of \textit{other-condemning} rhetoric interacts with reinforcement dynamics inherent to social media platforms. By criticizing perceived moral violations of others, such rhetoric distinguishes between in-groups and out-groups and may fuel polarization~\cite{brady2020mad, crockett2017moral, brady2017emotion, kim-etal-2024-moral}. When rewarded with engagement signals such as likes and comments, these expressions can become socially reinforced, mirroring a reinforcement-learning-like behavioral pattern~\cite{brady2021social}. As moral outrage becomes incentivized through these feedback signals, media outlets may increasingly adopt divisive framing strategies to sustain visibility within the attention economy.

While our analysis focuses on major news outlets, the amplification of moral outrage carries broader risks when exploited by malicious actors. Prior research shows that trolls and misinformation networks disproportionately employ \textit{other-condemning} language~\cite{simchon2022troll, solovev2022moral}. Given our finding that such rhetoric stimulates not only passive attention but also active engagement—through likes and comments that facilitate content diffusion—it can be readily exploited to amplify false or polarizing narratives across platforms.

Along with these concerns, we propose two directions for future work. One avenue is to investigate how platforms can reduce excessive moral outrage without undermining its constructive role in holding institutions accountable. Another is to incorporate \textit{other-condemning} rhetoric detection into troll and misinformation frameworks to enhance interpretability and diagnostic insight. Consistent with prior research~\cite{simchon2022troll, solovev2022moral,mcloughlin2024misinformation}, our findings suggest that measuring moral outrage offers higher-resolution lens for assessing the severity and societal impact of harmful content. To support further research and ensure reproducibility, we release our multimodal moral emotion classifiers for both Korean and English.
\section{Conclusion}
This study examined how moral emotions in multimodal media content shape user engagement across three behavioral levels (views, likes, and comments) using YouTube data from Korea and the United States. We found that \textit{other-condemning} rhetoric consistently increased engagement, with effects intensifying from viewing to commenting. These findings reveal that moral outrage is a powerful force in the attention economy, driving both attention and active participation. However, its effectiveness also entails risks: the divisive nature of \textit{other-condemning} rhetoric may deepen polarization when leveraged by media outlets or malicious actors.

Our contributions are twofold. Methodologically, we developed and publicly released the first multimodal moral emotion classifiers for both Korean and English contexts. Theoretically, we address a critical gap in cross-cultural moral emotion research by showing that \textit{other-condemning} rhetoric reliably elicits commitment-based engagement beyond mere attention across distinct cultural settings. Together, these contributions bridge moral psychology and computational social science, advancing our understanding of how moral emotions shape user behavior in algorithmically mediated environments.
\section{Limitations}
This study has several limitations that should be acknowledged. First, as an observational analysis, it identifies correlations between moral emotions in multimodal content and user engagement patterns without establishing causality. Future experimental work in controlled settings is needed to validate these findings and explore causal mechanisms. Second, while the U.S. annotation (Fleiss’ kappa=0.5271) achieved fair to good agreement—within the 0.40–0.75 range suggested by~\citet{fleiss2013statistical}—and exceeded prior benchmarks~\cite{kim-etal-2024-moral}, it showed lower reliability than the Korean dataset. This gap likely reflects presentation differences: Korean YouTube content frequently includes overlaid captions that make moral cues more explicit, as evidenced by the examples presented in Table~\ref{tab:moral_examples_split}. Future studies could leverage our released model to pre-select representative thumbnail–title pairs, which may particularly help improve annotation consistency for U.S. categories—\textit{self-conscious} and \textit{non-moral emotion}—where current performance remains limited. Third, although we used fine-tuning on human-labeled data to minimize reliance on biases in pre-trained LLMs (e.g., Western-centric tendencies~\cite{atari2023humans}), such biases may still persist. Users of our model should remain aware of these constraints. Despite these limitations, this work provides the first multimodal framework for measuring moral emotions and takes an important step toward broadening moral emotion research beyond Western-centric perspectives~\cite{van2024social}.

\begin{acks}
We sincerely thank Meeyoung Cha for her thoughtful feedback and helpful discussions throughout the development of this work. The anonymous reviewers are also gratefully acknowledged, along with our annotators and SoonHyeon Kwon, for their insightful discussions and support. This work was supported by the Hyundai Motor Chung MongKoo Foundation, the KAIST Cross-Generation Collaborative Lab Project (Project No. N11250080).

% , and the Ministry of Education of the Republic of Korea and the National Research Foundation of Korea (NRF-2025S1A5B5A20019277).
\end{acks}

% 한국연구재단: 석사과정생 연구장려금
% This work was supported by the Ministry of Education of the Republic of Korea and the National Research Foundation of Korea (NRF-2025S1A5B5A20019277)

% \pagebreak

% \newpage
\bibliographystyle{ACM-Reference-Format}
\bibliography{references}

% \newpage
\appendix
%\clearpage
\section{Appendix}
\subsection{Validation of Data Collection Time Point}
\label{sec:validation_data_collection}

\begin{figure}[th]
    \centering
    \includegraphics[width=0.86\linewidth]{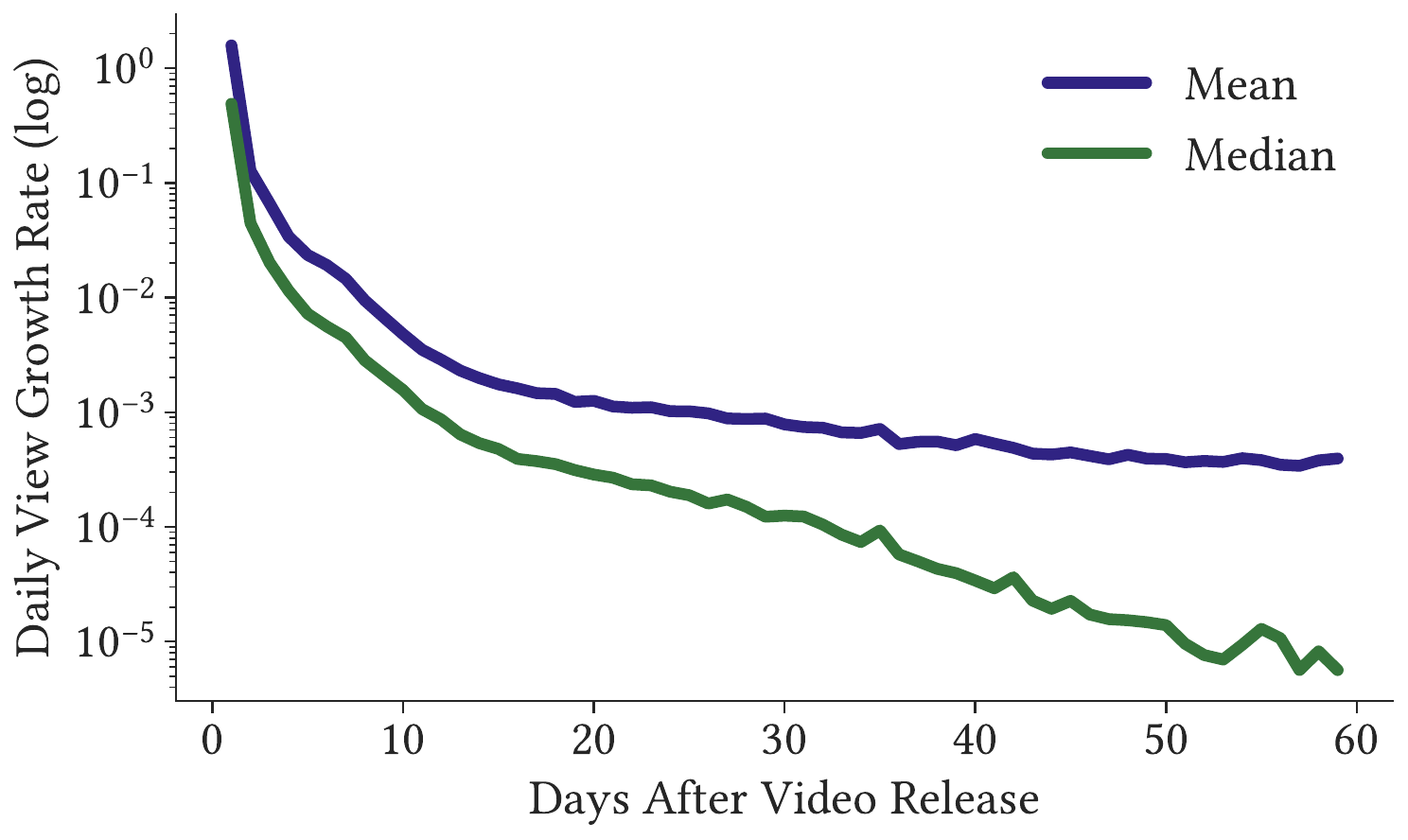}
    \caption{Mean and median daily view growth rates (log-scaled) across 1,703 videos over 60 days after release.}
    \label{fig:historical_stats}
    \vspace{-2mm}
\end{figure}

To verify engagement stability, we analyzed the daily view growth of 1,703 videos over a 60 days. The daily growth rate, calculated as the ratio of daily increases to the previous day's total, declines rapidly from an average of 0.6701\% in the first 10 days to 0.0395\% by day 60 (Figure~\ref{fig:historical_stats}). These findings confirm that engagement levels stabilize within two months, justifying our decision to standardize all metrics within a synchronized window (February 5–13, 2025). This approach ensures a fair comparison across channels while minimizing temporal bias from varying view accumulation periods.

% To verify when engagement metrics reach a stable point after video release, we analyzed the daily view growth dynamics of 1,703 selected samples from our dataset. For each video, view counts were collected once per day over a 60-day period following publication. The daily growth rate was computed as the ratio of the day-to-day increase in views to the previous day’s total views.

% Figure~\ref{fig:historical_stats} presents the log-scaled mean and median growth rates during the 60-day window. View growth declines rapidly during the initial period—averaging approximately 0.6701\% per day within the first 10 days, 0.0977\% by day 27, and 0.0395\% by day 60. These findings indicate that most news videos experience only marginal increases after the first month, confirming that engagement levels largely stabilize within two months of publication. This empirical validation supports our decision to standardize all engagement statistics within a synchronized collection window (February 5–13, 2025), ensuring fair comparison across all channels and videos while minimizing potential temporal bias due to differing view accumulation periods.

\subsection{Human Annotation Guideline and Interface}
\label{appendix:annotation_guideline}
Annotations for both Korean and English YouTube datasets were collected using POTATO, a web-based data annotation tool~\cite{pei2022potato}. 
% Figure~\ref{fig:annotation_guideline} displays the Korean and English guidelines provided to annotators. 
For the Korean dataset, six native annotators were recruited, each compensated with \textwon70,000 (totaling \textwon420,000). The English annotations were conducted in two rounds through Prolific\footnote{\url{https://www.prolific.com/}}, with three annotators per round. Total expenditure for English dataset, including platform fees, amounted to \textsterling164.00. All participants were provided with identical moral emotion category definitions, example cases, and interface instructions in their respective languages. The complete annotation guideline and interface screenshots are available at our GitHub repository.

\subsection{MLLM Prompt Format}
\label{appendix:llm prompts format}

Each prompt follows a chat format with three roles: system, user, and assistant. It includes a concise definition of the target moral emotion and a query asking whether the given thumbnail and title align with that emotion. Responses are restricted to \textit{True} or \textit{False}. For few-shot setting, the prompt includes one example that correctly reflects the target moral emotion, along with five additional examples that illustrate different, non-target moral emotions.

\subsection{Qwen vs LLaMA Performance Comparison}
\label{appendix:qwen_llama_comparison}

\begin{figure}[h!]
    \centering
    \includegraphics[width=0.86\linewidth]{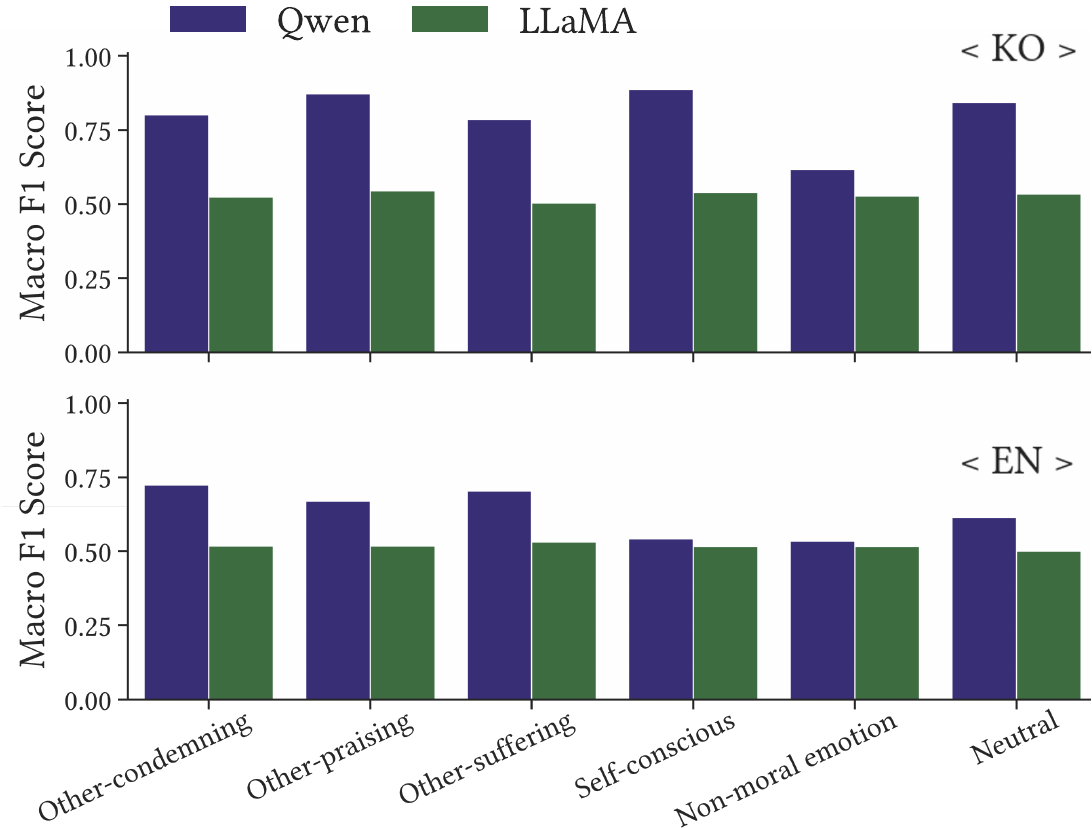}
    \caption{Comparison of macro F1 scores between the fine-tuned Qwen and LLaMA models across six moral emotion categories for both Korean (top) and English (bottom) datasets.}
    \label{fig:qwen_llama}
    \vspace{-1mm}
\end{figure}

To ensure robust open-source model selection, we compared \texttt{Qwen2\allowbreak-VL-7B-Instruct} and \texttt{Llama-3.2-11B-Vision-Instruct} under identical fine-tuning conditions. Qwen outperformed LLaMA, achieving average macro F1 scores of 0.8010 (Korean) and 0.6310 (English), compared to LLaMA's 0.5288 and 0.5168, respectively. Category-specific performance is visualized in Figure~\ref{fig:qwen_llama}. These results demonstrate Qwen's superior performance across both languages, motivating its selection as our final classifier.

% The results showed that while fine-tuning was the most effective approach for both models, Qwen consistently outperformed LLaMA. Specifically, the average macro F1 score for the fine-tuned Qwen model was 0.8010 for Korean and 0.6310 for English, compared to 0.5288 and 0.5168 for the LLaMA model, respectively. A detailed breakdown of the macro F1 scores for each of the six moral emotion categories is visualized in Figure~\ref{fig:qwen_llama}. These findings demonstrate Qwen’s superior performance across both languages, motivating its selection as the final classifier for our main analysis.

\onecolumn
\subsection{Robustness of Regression Analysis}
\label{sec:regression_robustness_analysis}

\begin{figure*}[bh!]
    \vspace{3mm}
    \centering
    \includegraphics[width=0.9\linewidth]{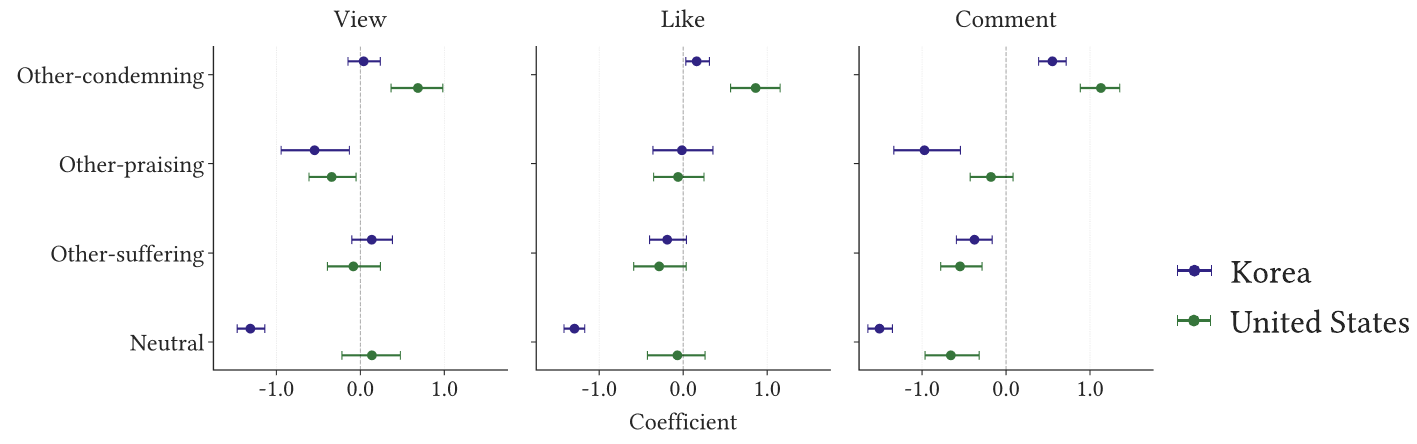}
    \caption{Bootstrap robustness check of regression coefficients. Points show mean estimates from 1,000 bootstrap regressions on random 20\% subsamples of the Korean and U.S. datasets, with horizontal lines indicating 95\% percentile confidence intervals. Panels show effects on view, like, and comment counts across four moral emotion categories.}
    \vspace{2mm}
    \label{fig:boot}
\end{figure*}

\begin{figure*}[bh!]
    \centering
    \includegraphics[width=0.9\linewidth]{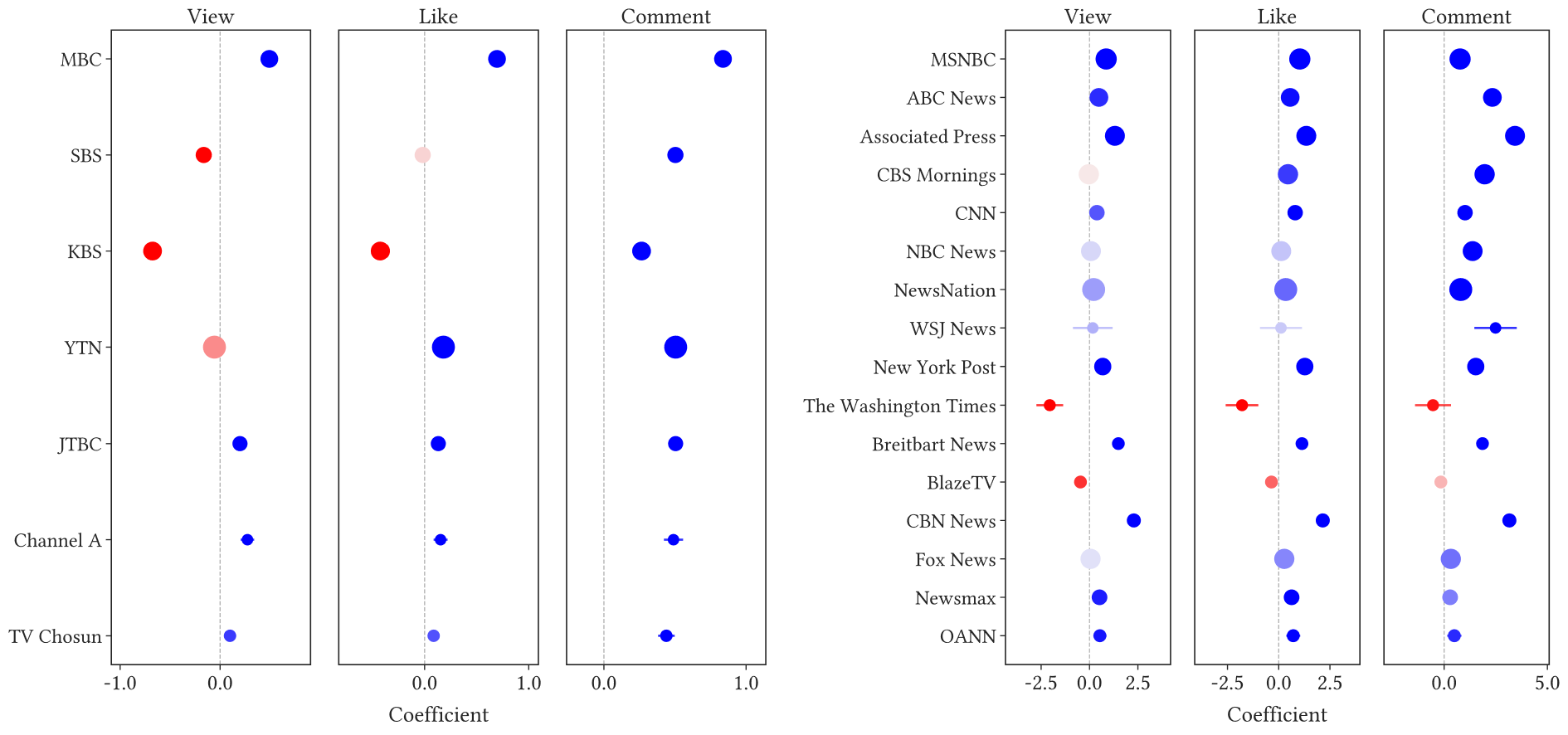}
    \caption{Channel-level regression coefficients for \textit{other-condemning} rhetoric. Points show estimated coefficients from channel-specific regression models predicting view, like, and comment counts; horizontal lines indicate 95\% confidence intervals. Color denotes coefficient sign (blue: positive; red: negative), and point size reflects the number of videos.}
    \label{fig:regression_by_channel}
\end{figure*}

To assess the stability of the main findings, we conducted two robustness analyses. First, bootstrap resampling was performed with 1,000 iterations using random 20\% subsamples. Figure~\ref{fig:boot} shows that the bootstrap estimates closely mirror the main results. Although some confidence intervals overlap zero due to increased uncertainty under subsampling, the overall directional patterns remain stable. Across both countries, \textit{other-condemning} rhetoric exhibits consistently positive mean effects across engagement outcomes, and the hierarchical pattern identified in the main analysis—where the effect strengthens from views to likes and comments—remains intact under repeated resampling. 
Second, separate regression models were estimated for each news channel in Korea and the United States to assess outlet-level heterogeneity. As shown in Figure~\ref{fig:regression_by_channel}, the channel-level estimates are largely consistent with the main regression results. Despite variation in effect magnitude, channel-level estimates indicate that the \textit{other-condemning} effect is positive for most outlets, with particularly strong associations for comments. This pattern persists across ideologically diverse media organizations.

\twocolumn

% In order to explore the best-performing model for our research, we experimented with another open-source MLLM model, which was Llama-3.2-11B-Vision-Instruct, and applied the same classification task fine-tuning approach as with Qwen ~\cite{llama3paper, llama3blog}. The hyperparameters were set identical to those used for Qwen. As illustrated in the table below, although Llama showed balanced performance across languages, Qwen consistently outperformed Llama in terms of both six moral emotions and two languages. Additionally, Qwen’s inference speed was over three times faster than Llama, ultimately leading to the selection of Qwen as the final model.

\begin{table*}[t]
\vspace{2mm}
  \centering
  \setlength{\tabcolsep}{6pt}
  \renewcommand{\arraystretch}{1.2}
  \caption{Overview of the Korean and U.S. YouTube news channels, including creation dates, subscriber counts, total uploaded videos, cumulative views, and videos in the 2024 dataset. Channels are ordered by political orientation within each country.}
  \begin{tabular}{c|cccc|c}
    \toprule
    \textbf{Channel} & \textbf{Creation Date} & \textbf{Subscribers} & \textbf{Total Videos} & \textbf{Total Views} & \textbf{Videos in Dataset (2024)} \\
    \midrule
MBC & 2006-11-05 & 5,990,000 & 340,262 & 24,797,909,341 & 47,081 \\ 
SBS & 2014-05-02 & 5,040,000 & 303,297 & 15,303,288,548 & 38,473 \\ 
KBS & 2013-08-06 & 3,450,000 & 402,625 & 9,438,028,884 & 53,030 \\ 
YTN & 2013-05-23 & 5,180,000 & 899,837 & 17,968,642,717 & 81,046 \\ 
JTBC & 2012-02-21 & 4,720,000 & 267,723 & 15,193,306,921 & 33,157 \\ 
Channel A & 2012-05-21 & 3,290,000 & 148,091 & 6,324,043,072 & 17,970 \\ 
TV Chosun & 2012-08-23 & 2,950,000 & 156,352 & 4,519,383,020 & 21,379 \\ \midrule 
MSNBC & 2011-12-01 & 9,170,000 & 103,524 & 17,802,177,849 & 12,824 \\ 
ABC News & 2006-08-07 & 19,000,000 & 105,389 & 17,194,229,268 & 8,858 \\ 
Associated Press & 2006-09-18 & 4,070,000 & 191,782 & 3,987,254,199 & 10,735 \\ 
CBS Mornings & 2013-05-23 & 3,210,000 & 48,283 & 2,646,175,644 & 11,281 \\ 
CNN & 2005-10-02 & 18,700,000 & 175,996 & 18,991,981,999 & 4,441 \\ 
NBC News & 2006-07-19 & 11,500,000 & 83,713 & 9,022,804,580 & 10,678 \\ 
NewsNation & 2020-05-06 & 2,380,000 & 75,204 & 1,500,720,042 & 15,861 \\ 
WSJ News & 2010-12-22 & 621,000 & 3,241 & 174,398,384 & 281 \\ 
New York Post & 2006-03-02 & 2,220,000 & 30,734 & 1,669,345,371 & 6,615 \\ 
The Washington Times & 2007-03-13 & 19,300 & 4,624 & 7,323,567 & 750 \\ 
Breitbart News & 2015-02-24 & 408,000 & 12,504 & 148,557,721 & 1,491 \\ 
BlazeTV & 2013-09-20 & 2,140,000 & 14,735 & 1,029,928,612 & 1,580 \\ 
CBN News & 2008-03-31 & 2,560,000 & 41,938 & 1,013,733,670 & 2,851 \\ 
Fox News & 2006-09-19 & 14,500,000 & 126,711 & 22,336,811,664 & 11,120 \\ 
Newsmax & 2008-01-18 & 2,460,000 & 53,921 & 843,288,176 & 4,722 \\ 
One America News Network (OANN) & 2013-10-04 & 1,410,000 & 17,723 & 228,071,971 & 1,673 \\ 
    \bottomrule
  \end{tabular}
  \label{tab:channels_info}
\end{table*}

\begin{table*}[bh!]
    % \vspace{6mm}
    \centering
    \footnotesize
    \renewcommand{\arraystretch}{1.6} 
    \caption{Example video thumbnails and titles from Korean and U.S. news channels for each moral emotion category.}
    \begin{adjustbox}{center}
    \begin{tabular}{c|c|c|c|c|c|c}
    \toprule
    & \textbf{Other-condemning} & \textbf{Other-praising} & \textbf{Other-suffering} & \textbf{Self-conscious} & \textbf{Neutral} & \textbf{Non-moral emotion} \\
    \midrule
    Korea
       & \begin{minipage}{0.14\textwidth}\centering
            \includegraphics[width=\linewidth]{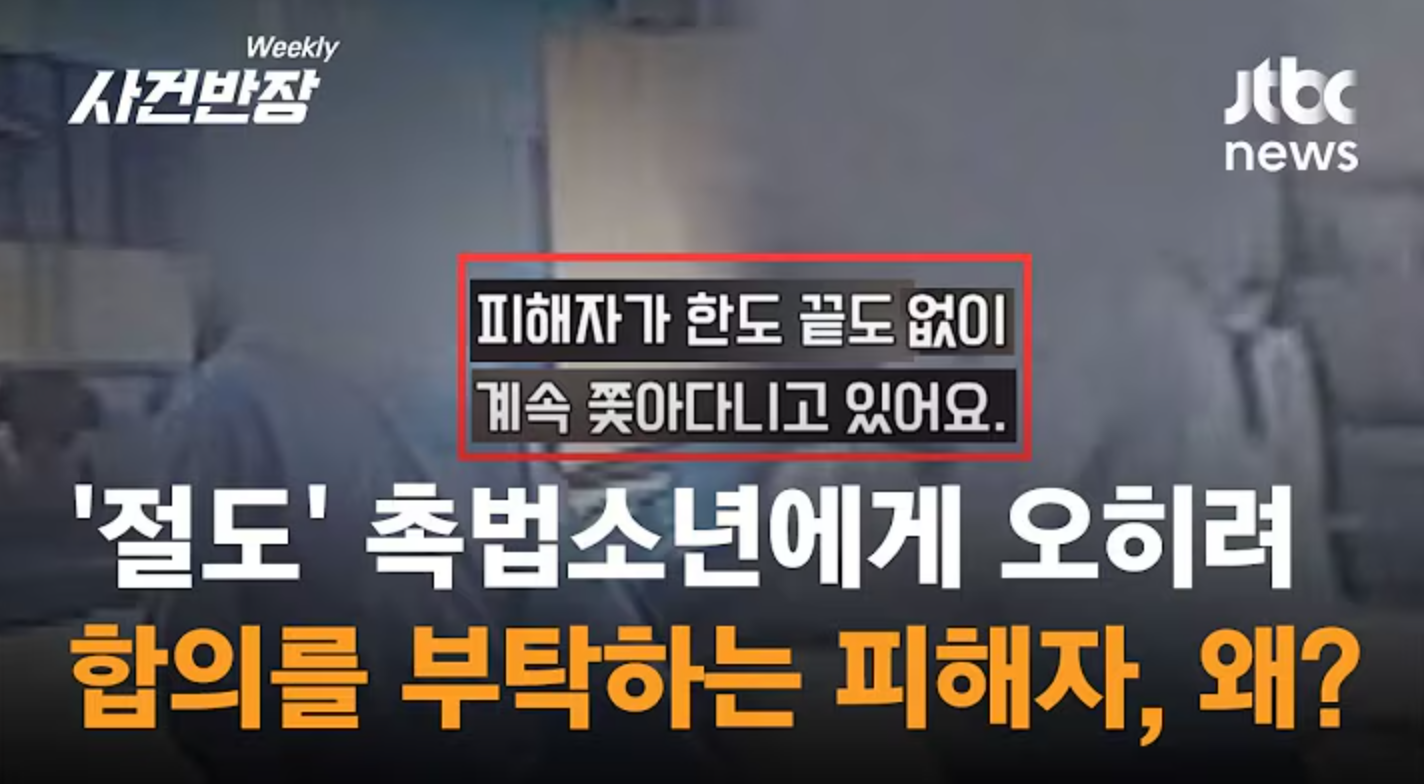}\\
            \scriptsize 오토바이 절도한 간 큰 \newline10대들… 사과·반성 없이 `뻔뻔'하게 하는 말이 \\
            (Brazen teens caught stealing motorcycles… Shameless remarks without apology or remorse)
         \end{minipage} 
       & \begin{minipage}{0.14\textwidth}\centering
            \includegraphics[width=\linewidth]{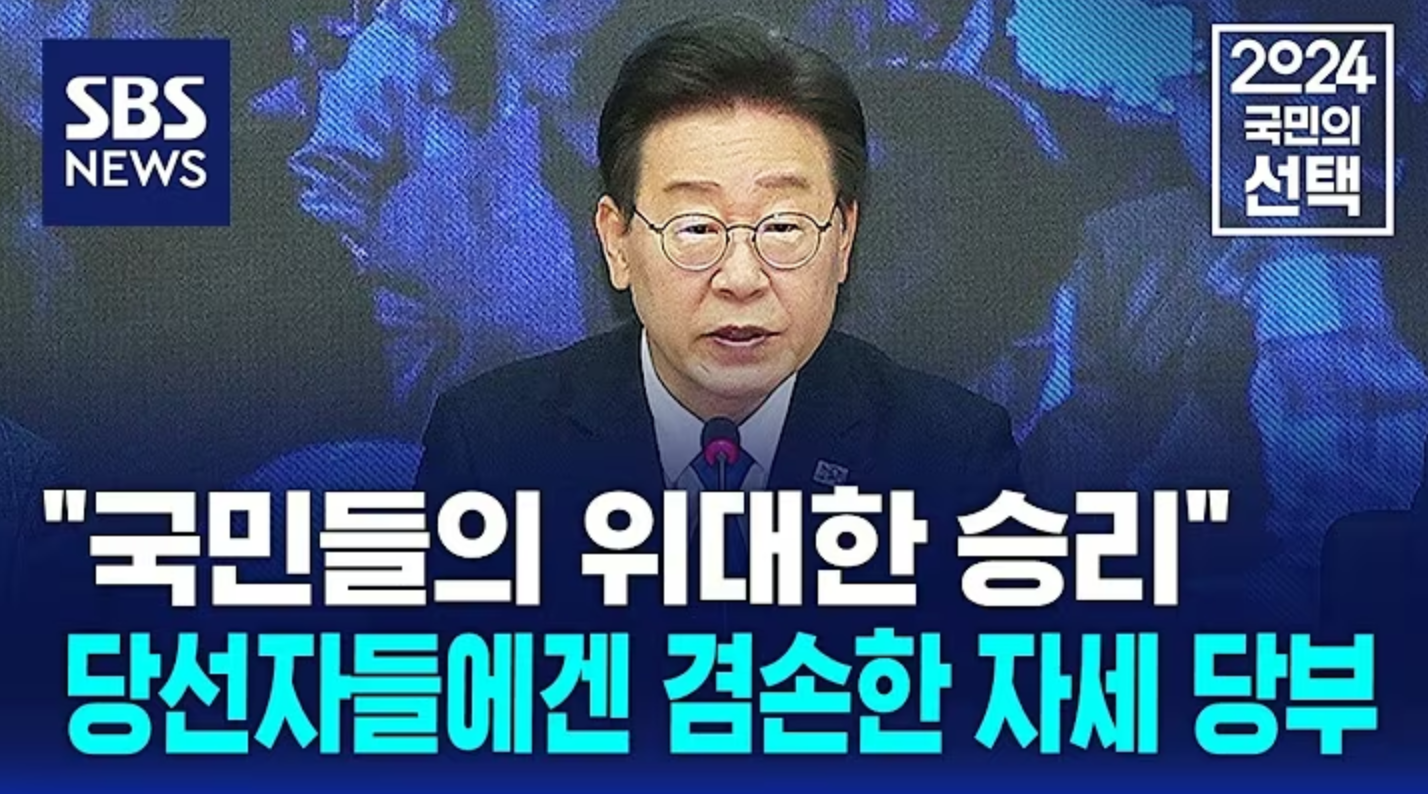}\\
            \scriptsize ``국민들의 위대한 승리”… 당선자들에겐 겸손한 \newline자세 당부 \\
            (``A great victory of the people”… Urging humility among the elected candidates)
         \end{minipage} 
       & \begin{minipage}{0.14\textwidth}\centering
            \includegraphics[width=\linewidth]{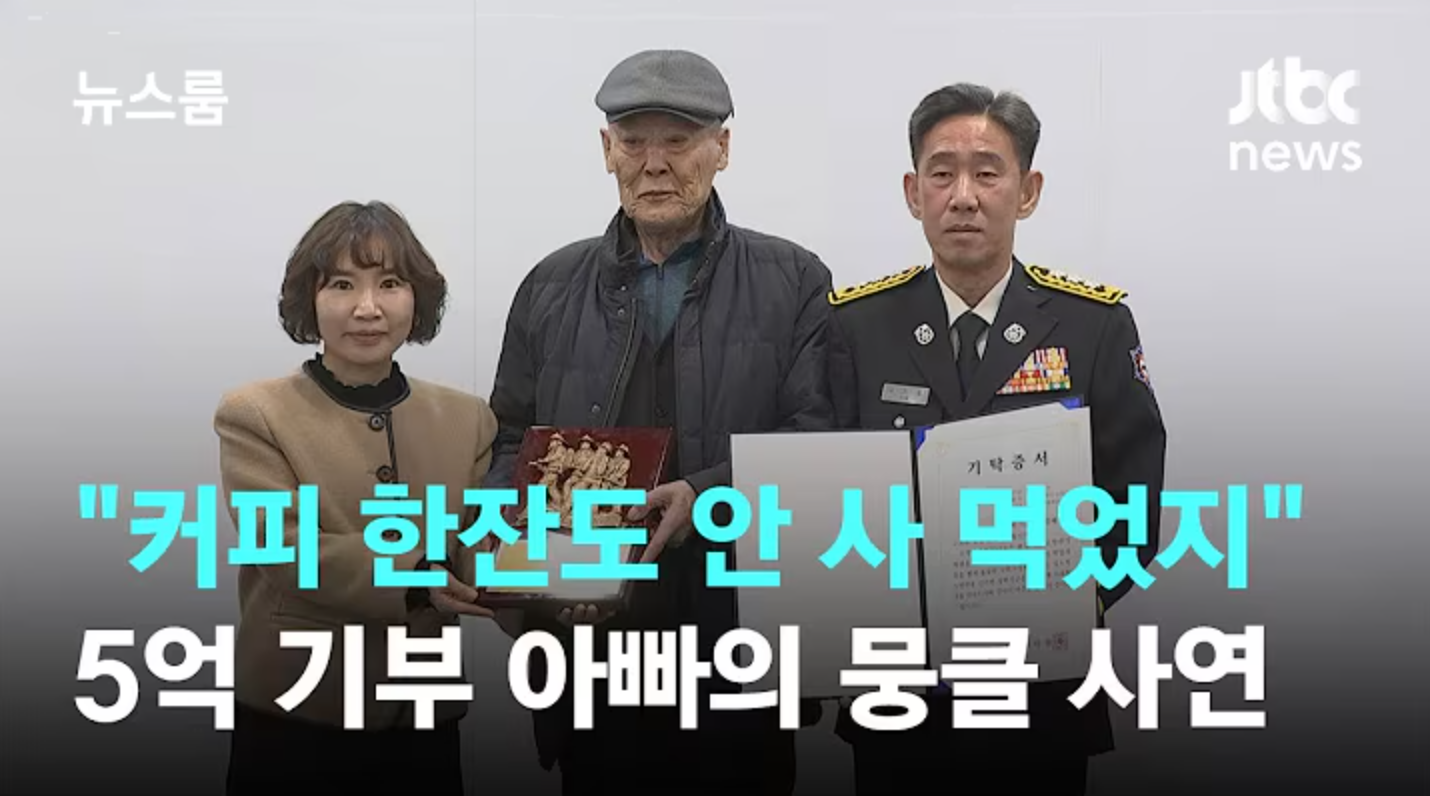}\\
            \scriptsize ``순직 소방관 자녀 위해 써주세요”… 아들 기리며 \newline`5억 장학금' \\
            (``Please use it for the children of fallen firefighters”... 500 million won scholarship in memory of his son)
         \end{minipage} 
       & \begin{minipage}{0.14\textwidth}\centering
            \includegraphics[width=\linewidth]{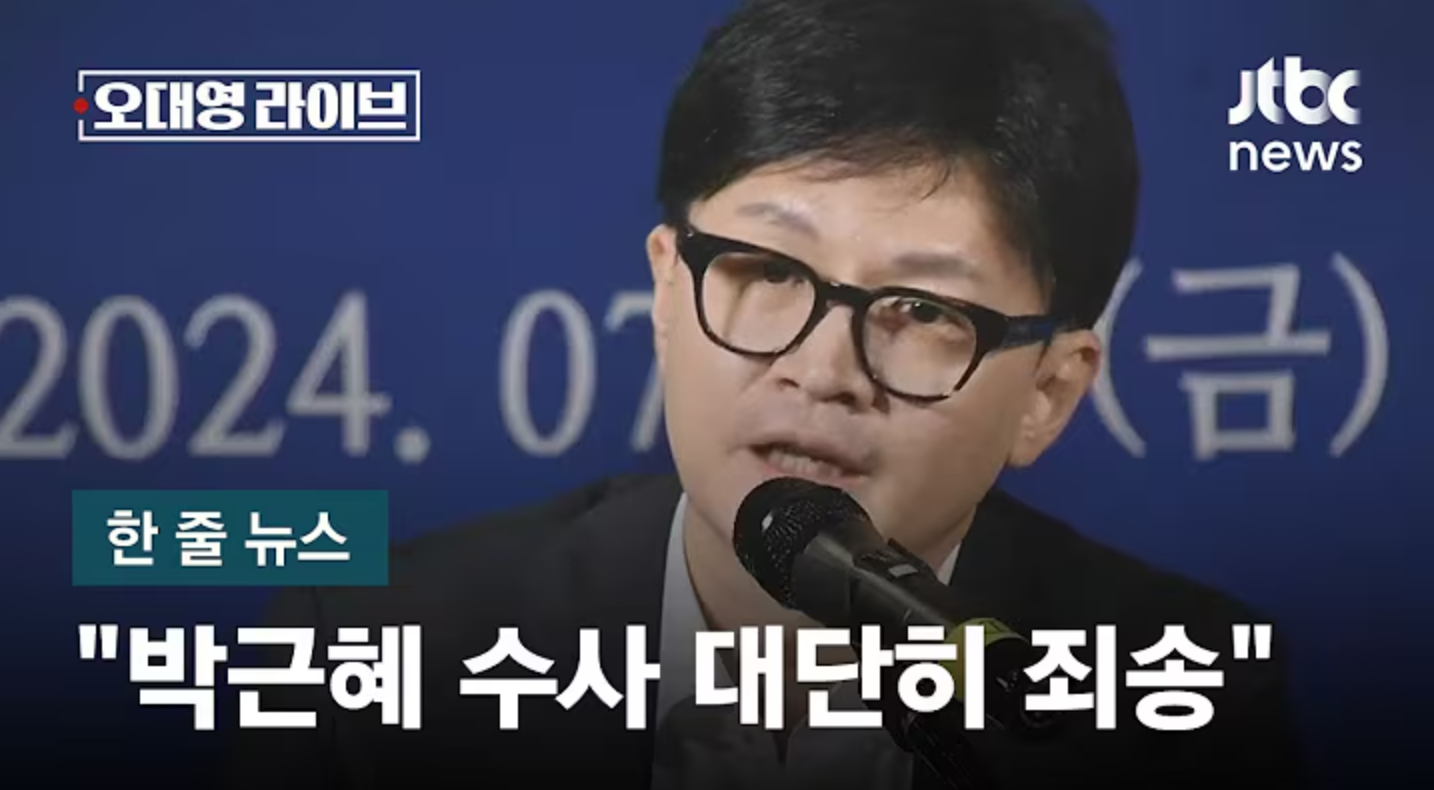}\\
            \scriptsize [한 줄 뉴스] 한동훈 ``박근혜 전 대통령 수사 대단히 죄송하게 생각해” \\
            ([One-line News] Han Dong-hoon: ``I deeply regret the investigation into former President Park Geun-hye”)
         \end{minipage} 
       & \begin{minipage}{0.14\textwidth}\centering
            \includegraphics[width=\linewidth]{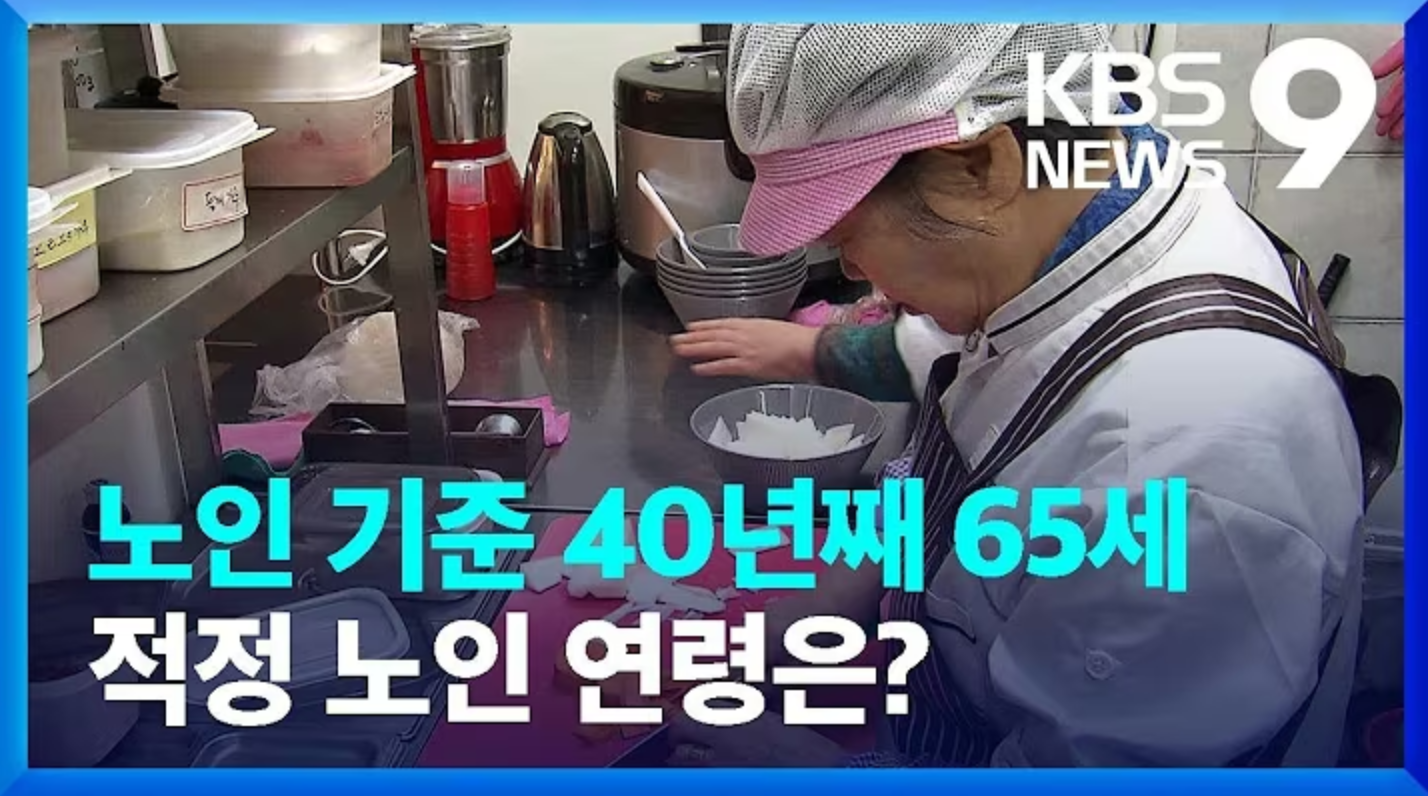}\\
            \scriptsize 40년째 노인은 `65세 이상’… 고령층 ``72세 이상이 노인” \\
            (For 40 years, seniors defined as `65+’… Elderly say ``72+ should be considered seniors”)
         \end{minipage} 
       & \begin{minipage}{0.14\textwidth}\centering
            \includegraphics[width=\linewidth]{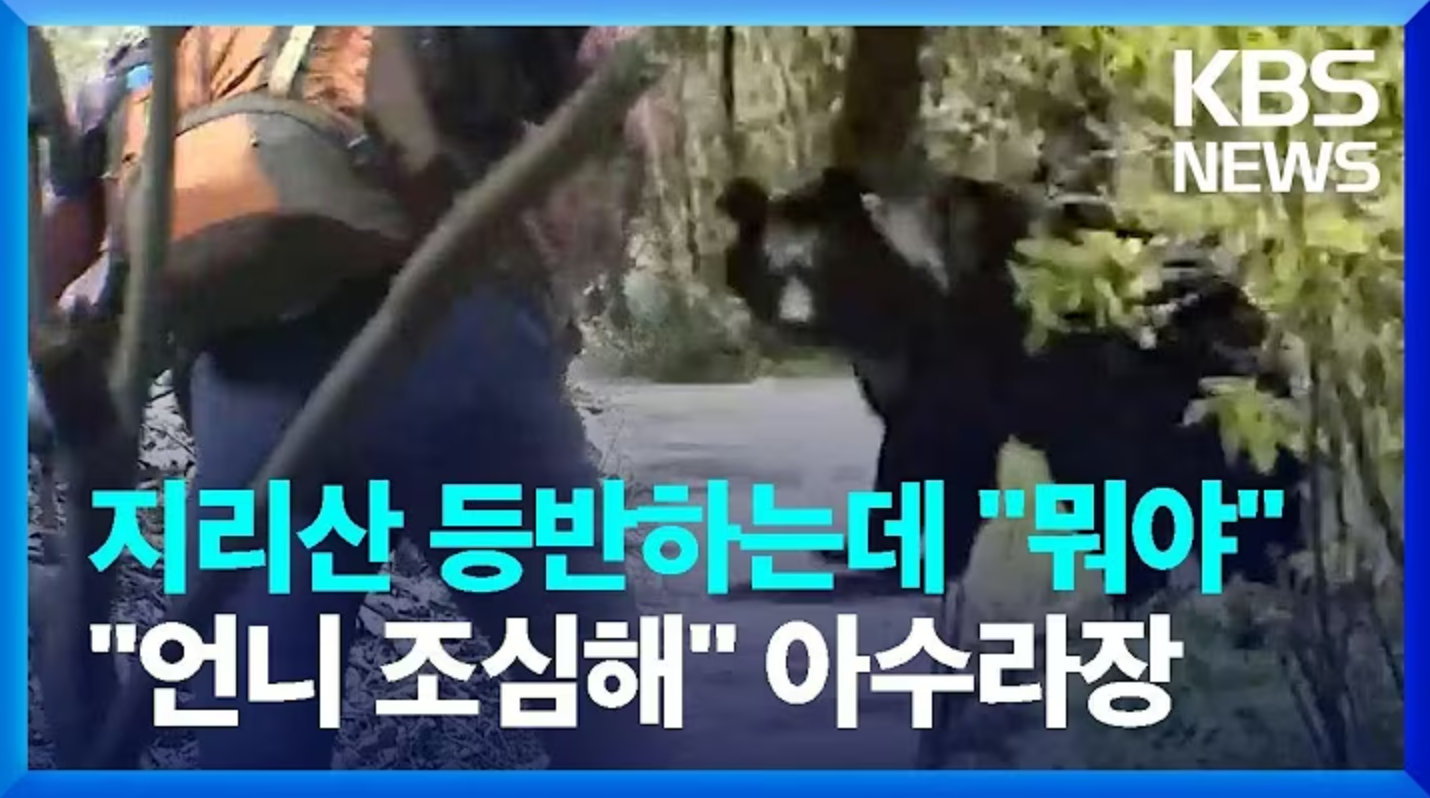}\\
            \scriptsize 지리산 등반하는데 ``뭐야” ``언니 조심해” 아수라장 \\
            (Climbing Jirisan turns chaotic with shouts of ``What’s going on?” and ``Sister, watch out”)
         \end{minipage} \\
    \midrule
    U.S.
       & \begin{minipage}{0.14\textwidth}\centering
            \includegraphics[width=\linewidth]{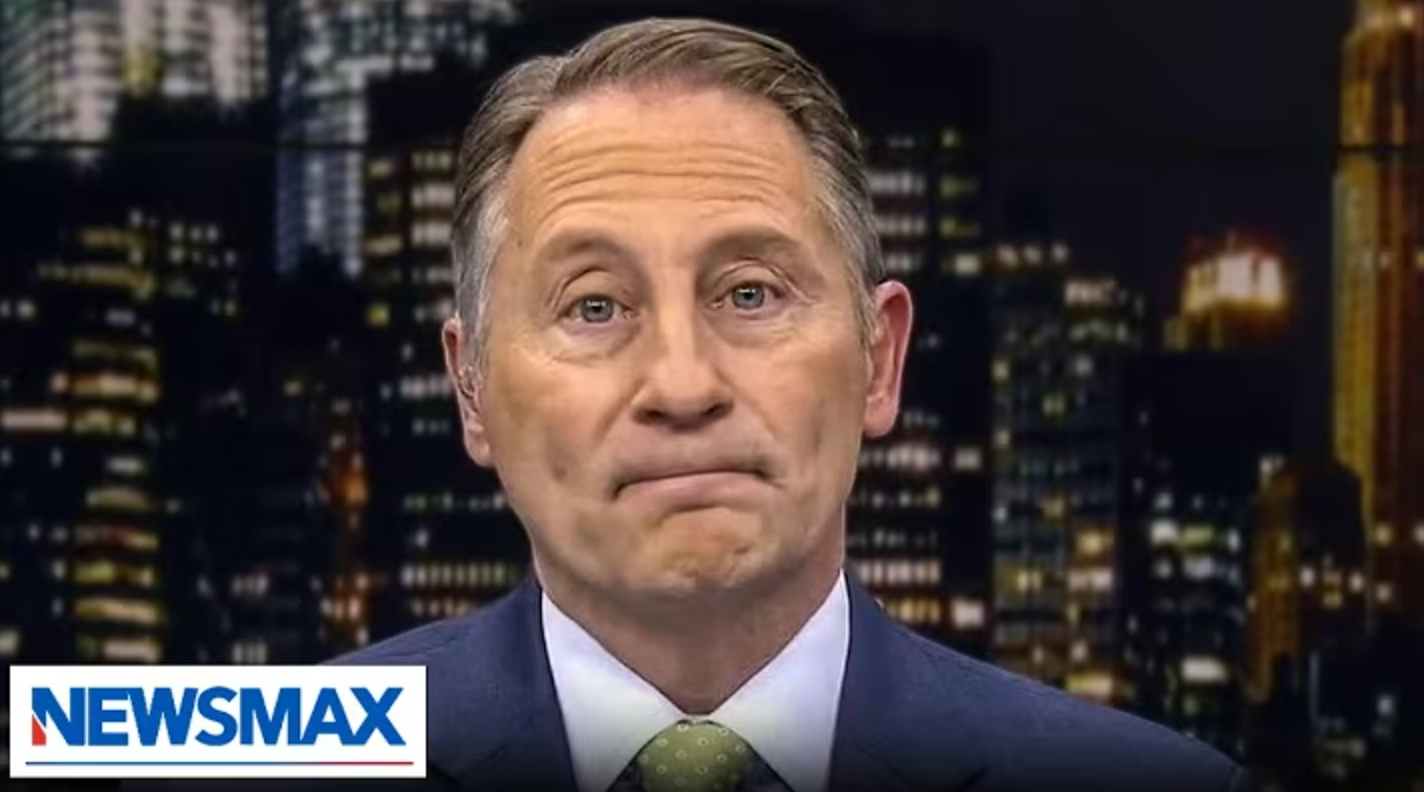}\\
            \scriptsize Part of `liberal elite’ mob still intent on keeping Trump out of White House: Rob Astorino
         \end{minipage} 
       & \begin{minipage}{0.14\textwidth}\centering
            \includegraphics[width=\linewidth]{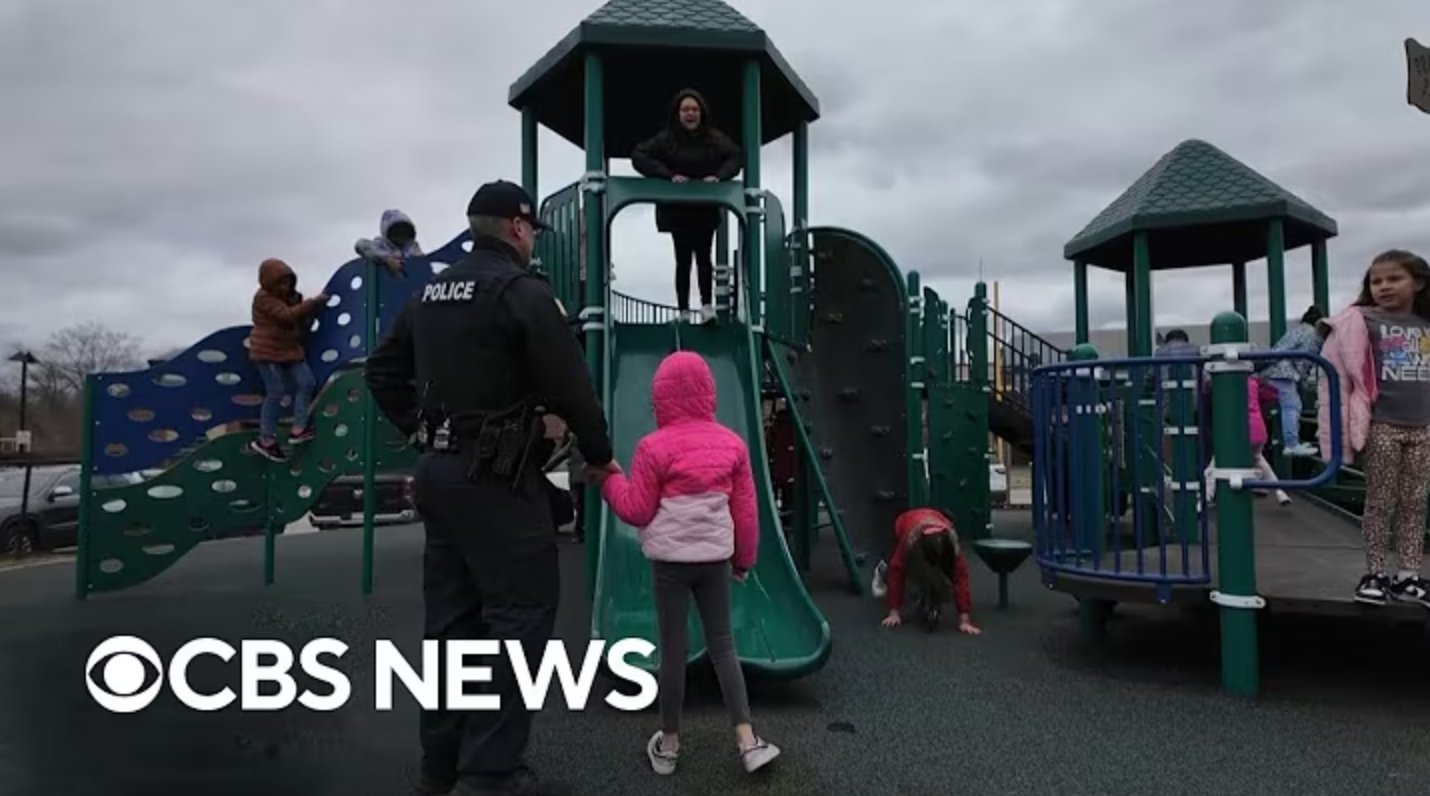}\\
            \scriptsize An angel officer | The Uplift
         \end{minipage} 
       & \begin{minipage}{0.14\textwidth}\centering
            \includegraphics[width=\linewidth]{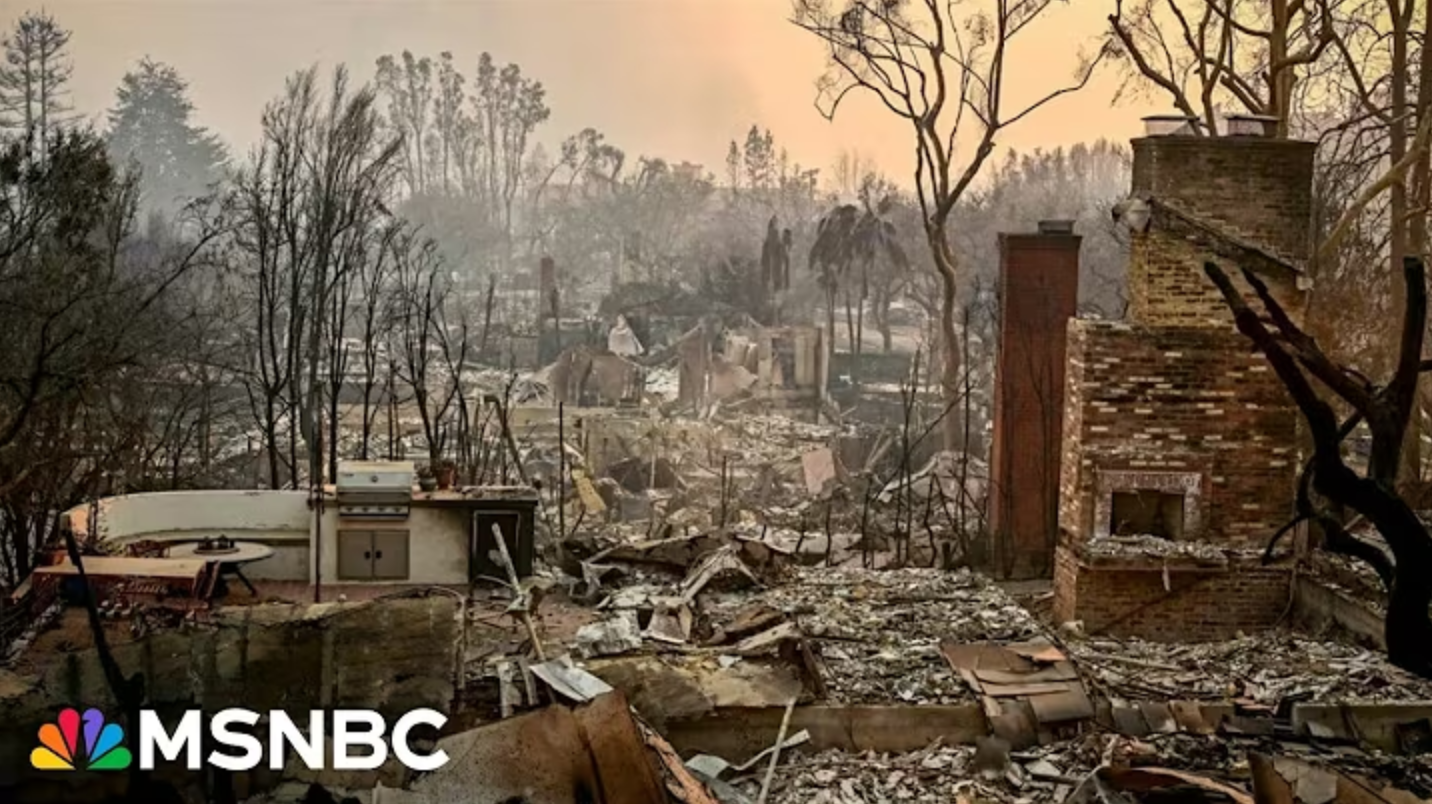}\\
            \scriptsize `Has to be hope' for recovery amid devastating aftermath of wildfires: Sen. Padilla
         \end{minipage} 
       & \begin{minipage}{0.14\textwidth}\centering
            \includegraphics[width=\linewidth]{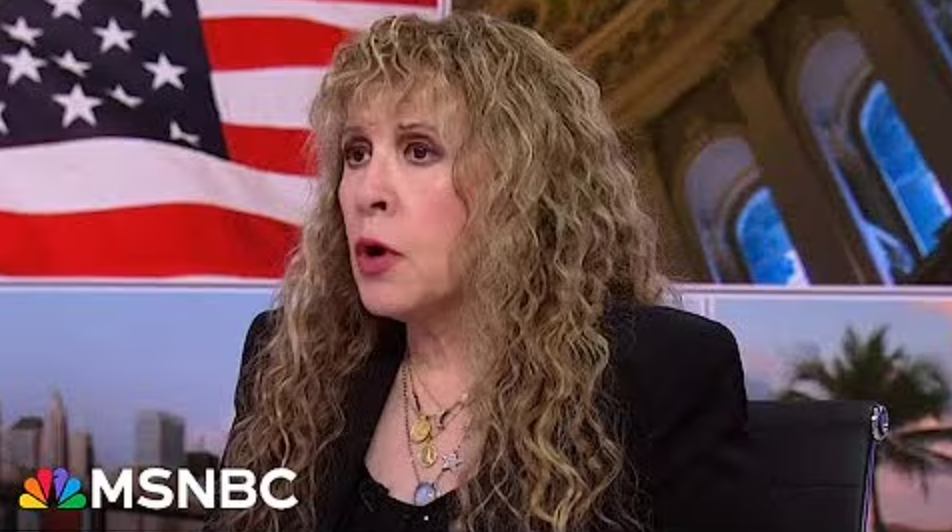}\\
            \scriptsize Stevie Nicks: I don't have many regrets, but I regret not voting until I was 70
         \end{minipage} 
       & \begin{minipage}{0.14\textwidth}\centering
            \includegraphics[width=\linewidth]{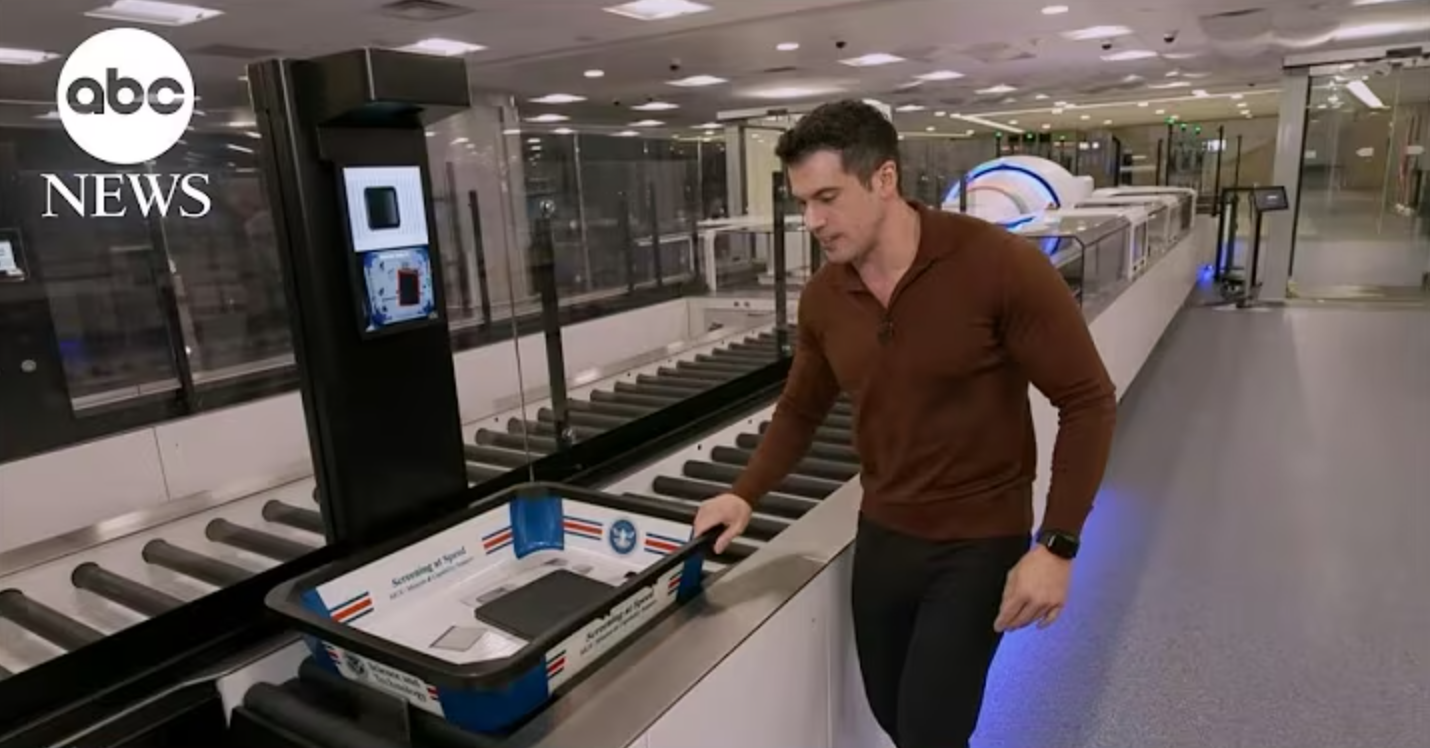}\\
            \scriptsize TSA tests self-checkout at Las Vegas airport
         \end{minipage} 
       & \begin{minipage}{0.14\textwidth}\centering
            \includegraphics[width=\linewidth]{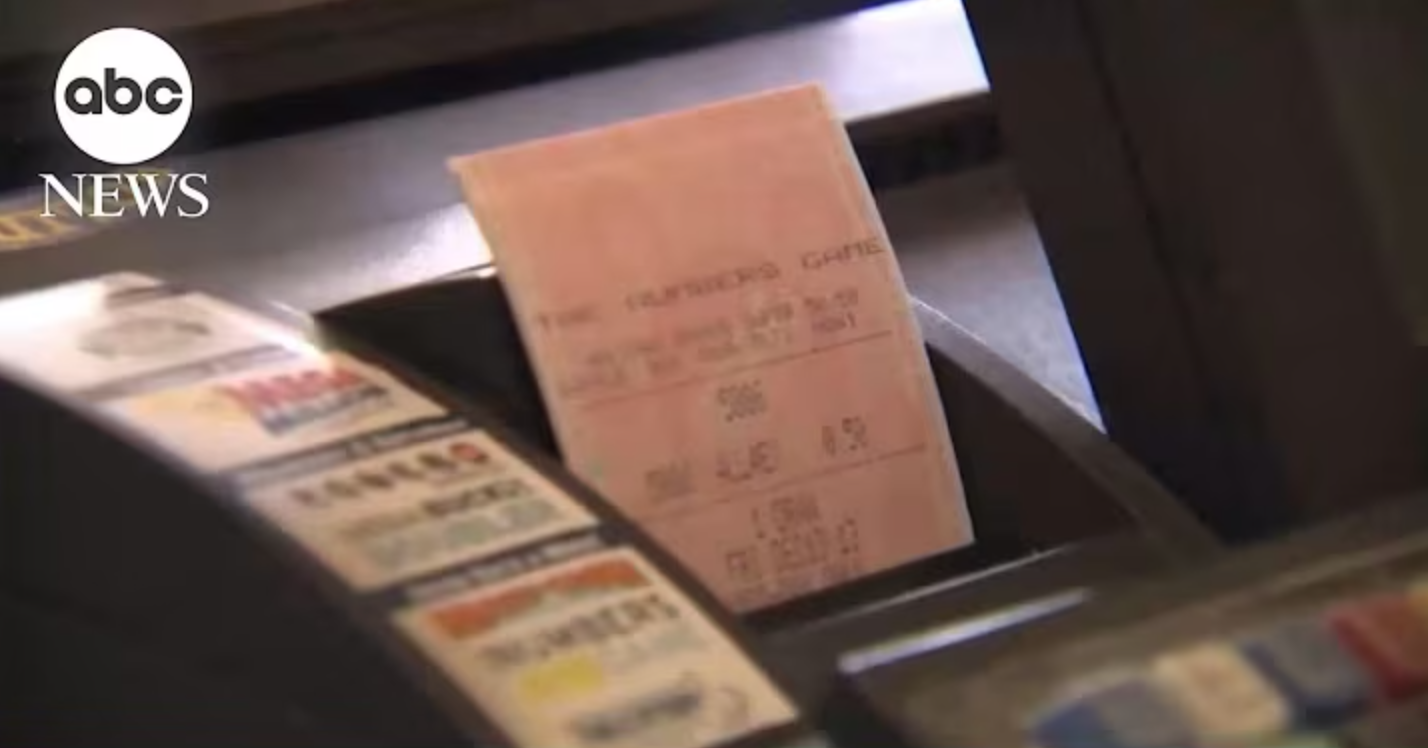}\\
            \scriptsize Americans feeling lucky for Christmas Eve’s \$1 billion Mega Millions lottery
         \end{minipage} \\
    \bottomrule
    \end{tabular}
    \end{adjustbox}
    \label{tab:moral_examples_split}
\end{table*}

% \twocolumn[
%     {\includegraphics[width=0.99\textwidth]{figures/annotation_guideline.pdf}
%     \vspace{-1.5mm}
%     \captionof{figure}{Moral emotion annotation guideline (top) and the web-based annotation interface used by annotators (bottom).}
%     \label{fig:annotation_guideline}
%     }
% ]

\begin{table*}[h!]
    \vspace{1mm}
    \centering
    \small
    \renewcommand{\arraystretch}{1.6}
    \caption{Overview of moral emotion categories with Korean and English descriptions.}
    \vspace{-0.5mm}
    \begin{adjustbox}{max width=\textwidth}
    \begin{tabular}{c|
        >{\centering\arraybackslash}m{0.35\linewidth}|
        >{\centering\arraybackslash}m{0.45\linewidth}}
    \toprule
    \textbf{} & \textbf{Korean} & \textbf{English} \\
    \midrule
    Other-condemning
       & 분노, 경멸, 혐오 등과 같이 타인을 비난하는 감정 
       & Emotions that condemn others, such as anger, contempt, or disgust. \\
    \midrule
    Other-praising
       & 감탄, 감사, 경외감 등과 같이 타인을 칭찬하는 감정 
       & Emotions that praise others, such as admiration, gratitude, or awe. \\
    \midrule
    Other-suffering
       & 연민, 동정 등과 같이 타인의 고통에 공감하는 감정 
       & \makecell[c]{Emotions of empathy for the suffering of others,\\such as compassion or sympathy.} \\
    \midrule
    Self-conscious
       & \makecell[c]{수치심, 죄책감, 당혹감 등과 같이\\자신을 부정적으로 평가하는 감정} 
       & \makecell[c]{Emotions that negatively evaluate oneself,\\such as shame, guilt, or embarrassment.} \\
    \midrule
    Neutral
       & 감정이 없거나 거의 없는 중립적인 카테고리 
       & A neutral category with no or few emotions. \\
    \midrule
    Non-moral emotion
       & 두려움, 놀라움, 기쁨, 낙관주의 등과 같이 감정은 있으나 다른 도덕감정 범주에 속하지 않는 감정 
       & \makecell[c]{Emotions that are not part of the other moral emotion categories,\\such as fear, surprise, joy, or optimism.} \\
    \bottomrule
    \end{tabular}
    \end{adjustbox}
    \label{tab:moral_emotion_descriptions}
\end{table*}

\begin{figure*}[h!]
\vspace{4.5mm}
\small
\centering
\begin{tcolorbox}[colback=gray!5!white, colframe=gray!100!black,  width=\linewidth]

\textbf{< Korean Prompt >} \\

\textbf{\textcolor{red}{System}}: 당신은 `도덕감정’ 분야에 대한 전문가입니다. 주어진 이미지와 텍스트에 나타나는 도덕감정이 \textbf{\{target emotion\}}인지 여부를 분류하세요. \textbf{\{target emotion\}}은 다음과 같이 정의할 수 있습니다: \textbf{\{target emotion\}}에 대한 설명. \\

\textbf{\textcolor{blue}{User}}: 주어진 Thumbnail image와 Title text에 모두 나타난 도덕감정이 \textbf{\{target emotion\}}이 맞나요? 추가적인 설명 없이, 도덕감정이 \textbf{\{target emotion\}}인지 여부만 \textit{True} 또는 \textit{False}로 답변해주세요. \\
            
\textbf{Title}: ``순직 소방관 자녀 위해 써주세요”… 아들 기리며 `5억 장학금'
 
\textbf{Image}: 

\begin{center}
    \includegraphics[width=0.3\linewidth]{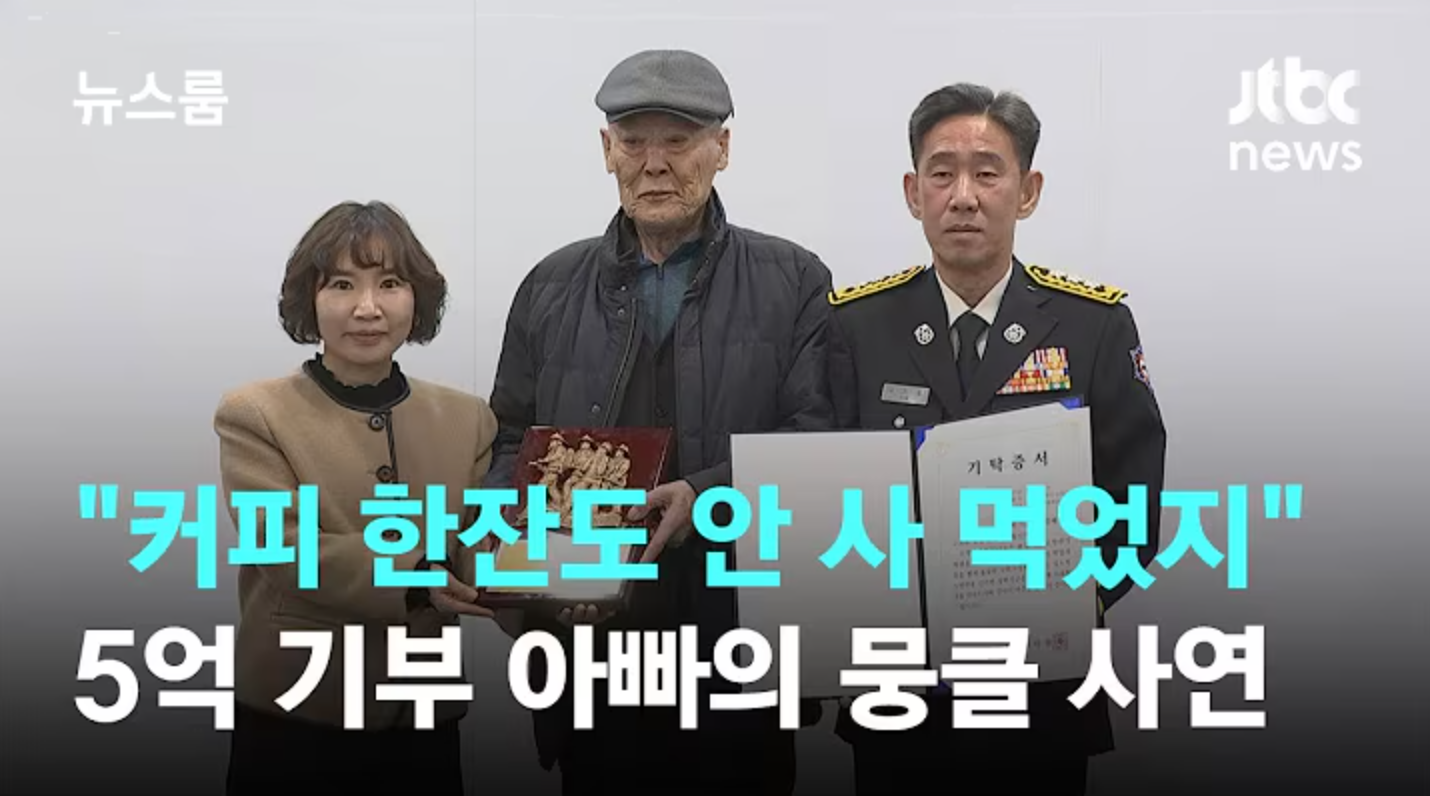}   
    \scriptsize
\end{center} 

\textbf{< English Prompt >} \\

\textbf{\textcolor{red}{System}}: You are an AI expert in `moral emotions'. Classify whether the moral emotion expressed in the given image and text is \textbf{\{target emotion\}}. \textbf{\{target emotion\}} is defined as follows: a description of \textbf{\{target emotion\}}.\\

\textbf{\textcolor{blue}{User}}: Is the moral emotion expressed in both the given Thumbnail image and Title text \textbf{\{target emotion\}}?
Without further explanation, answer only \textit{True} or \textit{False} to indicate whether the moral emotion is \textbf{\{target emotion\}}. \\

\textbf{Title}:  `Has to be hope' for recovery amid devastating aftermath of wildfires: Sen. Padilla

\textbf{Image}: 

\begin{center}
    \includegraphics[width=0.3\linewidth]{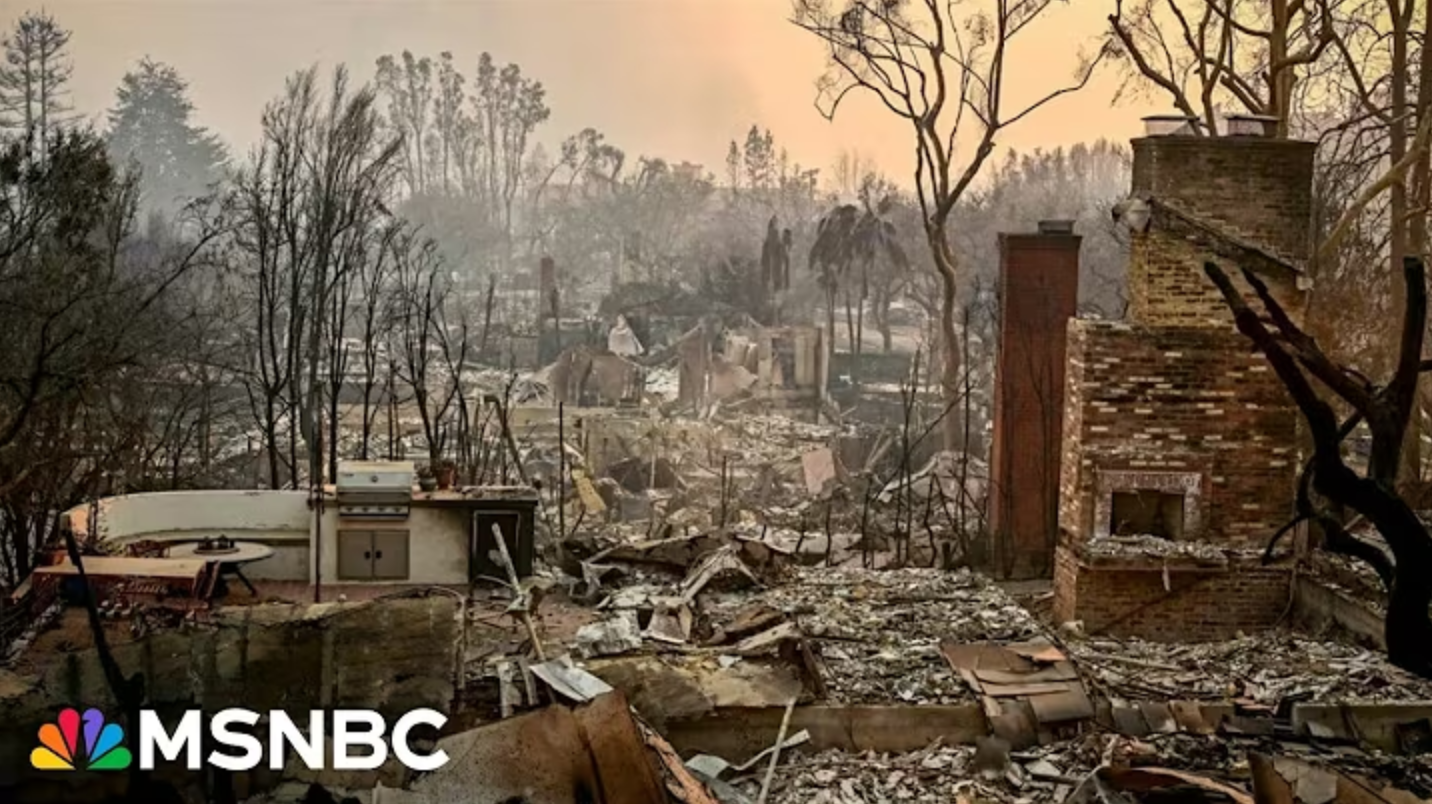}    
    \scriptsize
\end{center}
\end{tcolorbox}
\vspace{-0.5mm}
\caption{Prompting template used for moral emotion classification across Korean and English datasets.}
\label{fig:prompt_structure}
\end{figure*}

\clearpage

\end{document}